\documentclass[twocolumn,twocolappendix,trackchanges]{aastex701}

\usepackage{amsmath,amssymb}
\usepackage{physics}
\usepackage{amsfonts}
\usepackage{mathtools}
\usepackage{footmisc}

\usepackage{etoolbox}
\usepackage[refpage]{nomencl}

\providetoggle{nomsort}
\settoggle{nomsort}{true} 

\makeatletter
\iftoggle{nomsort}{%
    \let\old@@@nomenclature=\@@@nomenclature        
        \newcounter{@nomcount} \setcounter{@nomcount}{0}%
        \newcommand{\threedigits}[1]{\ifnum#1<100 0\two@digits{#1} \else \number#1\fi}
        \renewcommand\the@nomcount{\threedigits{\value{@nomcount}}}
        \def\@@@nomenclature[#1]#2#3{
          \addtocounter{@nomcount}{1}%
        \def\@tempa{#2}\def\@tempb{#3}%
          \protected@write\@nomenclaturefile{}%
          {\string\nomenclatureentry{\the@nomcount\nom@verb\@tempa @[{\nom@verb\@tempa}]%
          \begingroup\nom@verb\@tempb\protect\nomeqref{\theequation}%
          |nompageref}{\thepage}}%
          \endgroup
          \@esphack}%
      }{}
\makeatother

\newcommand{\mynomone}[3][section]{%
  \begingroup\edef\x{\endgroup
  \unexpanded{\nomenclature{#2}}%
    {\unexpanded{#3} \hspace*{\fill}  (\csname the#1\endcsname)}}\x}

\newcommand{\mynomtwo}[4][section]{%
  \begingroup\edef\x{\endgroup
  \unexpanded{\nomenclature[#2]{#3}}%
    {\unexpanded{#4} \hspace*{\fill}  (\csname the#1\endcsname)}}\x}

\renewcommand\nomgroup[1]{%
  \item[\bfseries
  \ifstrequal{#1}{A}{Acronyms}{%
  \ifstrequal{#1}{S}{Symbols}{%
  \ifstrequal{#1}{C}{Other Symbols}{}}}%
]}

\newcounter{logglabel}

\makenomenclature

\newcommand{\eq}[1]{Eq.~(\ref{#1})}

\shorttitle{Dynamical Lava Tides on Exoplanets}
\shortauthors{Farhat \& Chiang}
\graphicspath{{./}{figures/}}

\begin{document}

\title{Magma Ocean Waves and Thermal Variability on Lava Worlds}

\author[0000-0001-7864-6627]{Mohammad Farhat}
\altaffiliation{Miller Fellow}
\email{mfarhat@berkeley.edu}

\affiliation{Department of Astronomy, University of California, Berkeley, Berkeley, CA 94720-3411, USA  }
\affiliation{Department of Earth and Planetary Science, University of California, Berkeley, Berkeley, CA 94720-4767, USA}

\author[0000-0002-6246-2310]{Eugene Chiang}
\email{echiang@astro.berkeley.edu}
\affiliation{Department of Astronomy, University of California, Berkeley, Berkeley, CA 94720-3411, USA  }
\affiliation{Department of Earth and Planetary Science, University of California, Berkeley, Berkeley, CA 94720-4767, USA}



\begin{abstract}
Lava worlds are rocky planets with dayside skins made molten by stellar irradiation. Tidal heating on these shortest-period planets is more than skin deep. We show how orbital eccentricities of just a few percent (within current observed bounds and  maintained secularly by exterior companions) can create deep magma oceans. ``Lava tidal waves'' slosh across these oceans; we compute the multi-modal response of the ocean to tidal forcing, subject to a coastline at the day-night terminator and a parameterized viscous drag. Wave interference produces a dayside heat map that is spatially irregular and highly time-variable; hotspots can wander both east and west of the substellar point, and thermal light curves can vary and spike aperiodically, from orbit to orbit and within an orbit. Heat deposited by tides is removed in steady state by a combination of fluid, mushy, and solid-state convection in the mantle. For Earth-sized planets with sub-day periods, the entire mantle may be tidally liquified.

\end{abstract}



\section{Introduction} 
\label{sec:intro}

While some of the shortest period $(P_{\rm orb} < 10~\rm d)$ rocky planets may have had their dayside surfaces melted by stellar irradiation \citep[e.g.][]{kite2016atmosphere,nguyen2020modelling,lai2024ocean}, the degree to which such ``lava worlds'' are tidally heated is unclear. At least in principle, by analogy with the Solar System satellites Io, Europa, and Enceladus 
\citep[e.g.][]{moore2003tidal,sotin2009tides,renaud2018increased,meyer2007tidal}, tidal heating could dominate the energy budgets of the closest-in exoplanets, liquifying large portions of their interiors {\citep[e.g.][]{leger2011extreme,boukare2022deep,boukare2025role}.}

A key parameter is orbital eccentricity, which only needs to be on the order of a percent for tidal dissipation to be significant \citep[e.g.][]{bolmont2013tidal,ferraz2025tidal}. 
Of the 158 $P_{\rm orb} \leq 10~{\rm d}$ planets in the NASA Exoplanet Archive that are presumably rocky (here defined as having masses $M_{\rm p} < 10 M_\oplus$ and radii $R_{\rm p} < 2 R_{\oplus}$), 56 have reported eccentricities $e_{\rm p}$ of a few percent, consistent with zero to within 1-2$\sigma$; note that published eccentricities, especially those near the noise limit, are overestimated insofar as they do not account for the bias in measuring positive-definite quantities \citep[e.g.][]{zakamska2011observational}. Most of the remaining short-period rocky planets list either zero or no eccentricity at all. There are a few exceptions of higher formal significance, including Ross~176b ($P_{\rm orb} = 5$ days, $M_{\rm p} = 4.57^{+0.89}_{-0.93} \, M_\oplus$, $R_{\rm p} = 1.84\pm 0.08 \, R_\oplus$) with $e_{\rm p}=0.25\pm 0.04$ (\citealt{geraldia2025discovery}); however, as these authors note, the radial-velocity curve for Ross 176 from which $e_{\rm p}$ is determined does not sample all orbital phases. Another marginal example is K2-38b ($P_{\rm orb} = 4.0$ d, $M_{\rm p} = 7.3^{+1.1}_{-1.0} \, M_\oplus$, $R_{\rm p} = 1.54\pm 0.14 \, R_\oplus$) for which $e_{\rm p}=0.197^{+0.067}_{-0.060}$ \citep[][]{toledo2020characterization}.

Even if the eccentricities of lava worlds can be shown convincingly to be non-zero, there remains the question of how to maintain such eccentricities against tidal circularization over system ages which are typically several Gyrs. Tidally heated Solar System satellites solve this problem by being continuously gravitationally forced in mean-motion resonance with exterior satellites \citep[e.g.][]{peale1976orbital}. The same solution can apply in principle to extrasolar short-period planets; see e.g.~K2-138b ($P_{\rm orb} = 2.35$ days, $e_{\rm p} < 0.2$), the innermost planet of a mean-motion resonant chain \citep{christiansen2018k2}.

\begin{figure}[t]
\includegraphics[width=.47\textwidth]{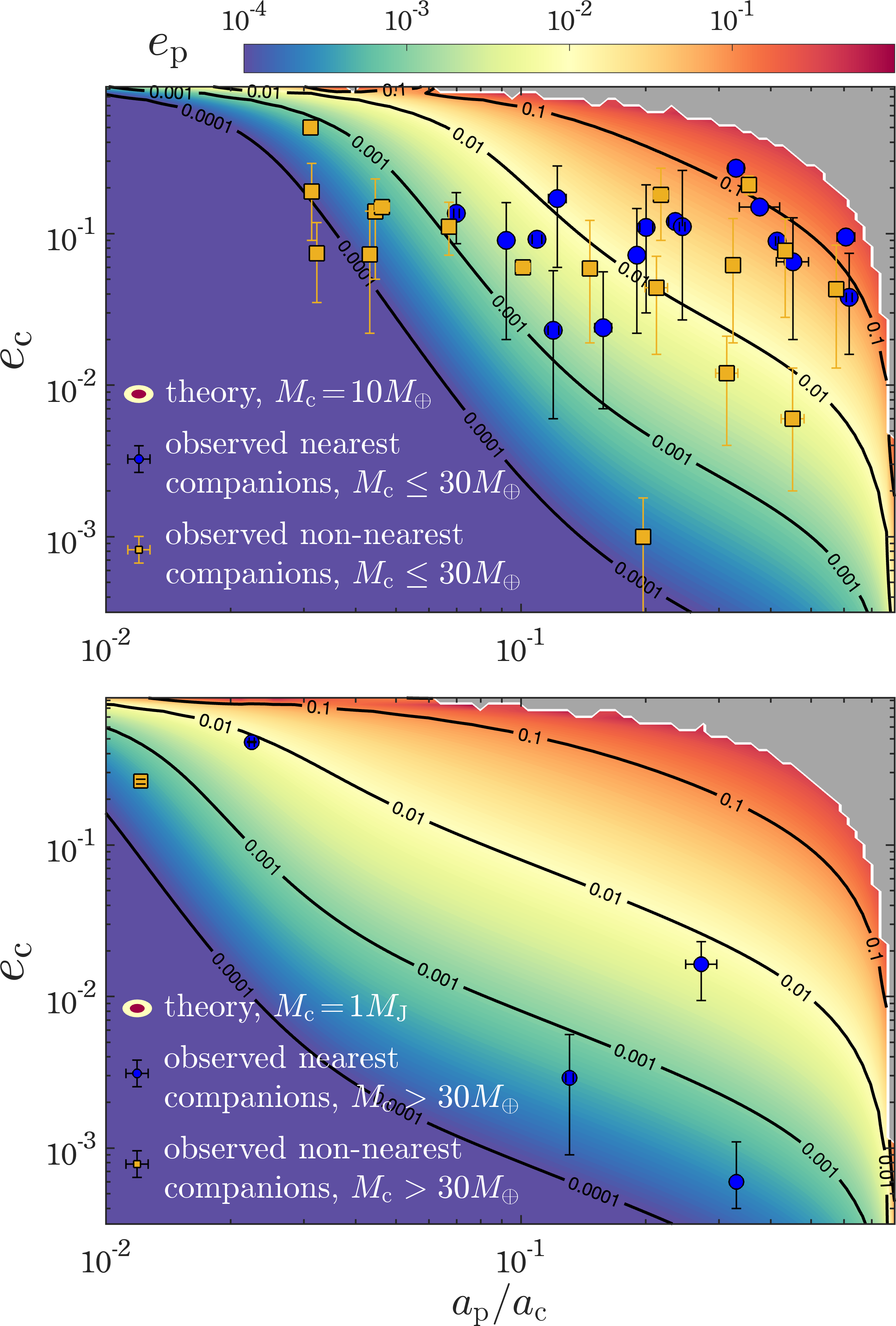}
\caption{Secular forcing of a short-period rocky planet's eccentricity by an outer companion. Colored contours plot solutions to \eq{forced_ecc_general} for the forced eccentricity $e_{\rm p}$, as a function of the outer companion's eccentricity $e_{\rm c}$ and the inner-to-outer semimajor axis ratio $a_{\rm p}/a_{\rm c}$. These contours assume the inner planet is a super-Earth of mass $M_{\rm p}=8M_{\oplus}$ on a 3-day orbit, while the outer planet has a mass $M_{\rm c}$ of either $10M_\oplus$ (upper panel) or $1M_{\rm J}$ (lower panel). The gray area at top right marks the region where inner and outer orbits cross and our (octopole-level) secular theory is invalid. Overlaid on the contours are data for known outer companions ($M_{\rm c}\leq 30M_\oplus$ in the upper panel,  $M_{\rm c}>30M_\oplus$ in the lower panel) to known rocky planets with $P_{\rm orb}\leq 3~$days. These observational data in combination with the theoretical contours suggest that a subset of short-period rocky planets may have $e_{\rm p}> 0.01$ and are significantly heated by eccentricity tides. For details, including calculations for how long $e_{\rm p}$ and $e_{\rm c}$ can be maintained against tidal dissipation, see Appendix~\ref{Section_maintaining_eccentricity}.  }
\label{Fig_ep_sf}
\end{figure}

Eccentricities can also be secularly forced by non-resonant eccentric companions, which many short-period rocky planets have. Using the same NASA Exoplanet Archive sample of  rocky worlds defined above, the fraction of $P_{\rm orb}\leq 10~{\rm d}$ planets known to have outer companions is 67\%. For $P_{\rm orb} \leq 1~{\rm d}$, 
the known companion fraction is 60\%. Figure~\ref{Fig_ep_sf} shows, for planets with $P_{\rm orb} \leq 3~{\rm d}$, published values for companion eccentricities $e_{\rm c}$, plotted against companion proximity as measured by the inner-to-outer semimajor axis ratio $a_{\rm p}/a_{\rm c}$. The larger are $e_{\rm c}$ and $a_{\rm p}/a_{\rm c}$ (the closer the companion is to the planet), the larger the eccentricity $e_{\rm p}$ that can be secularly forced. Overlaid on these data are theoretical contours for the forced $e_{\rm p}$ from a hypothetical companion of fixed mass $M_{\rm c}$, either $10 M_\oplus$ (upper panel) or $1 M_{\rm J}$ (lower panel). Figure \ref{Fig_ep_sf} suggests that for a subset of systems, forced eccentricities $e_{\rm p} > 0.01$ are possible.

The secular theory underlying Figure \ref{Fig_ep_sf} is detailed in Appendix~\ref{Section_maintaining_eccentricity}, where we also show how percent-level eccentricities can be sustained over Gyr timescales against tidal dissipation. We supply these calculations to argue that eccentricities of a few percent are plausible for some (not all) short-period rocky planets. Making the case for eccentricity --- or other sources of 
time-dependent tidal strain, such as obliquity or asynchronous rotation 
--- is not, however, the focus of this paper. Our goal instead is to explore what the thermodynamic consequences of tidal heating are for lava worlds. Can tidal stresses be so large that, instead of substellar magma ``ponds'' only $\sim$100 m deep \citep{kite2016atmosphere}, the entire dayside is covered by a magma ocean penetrating into the mantle, perhaps even to the metallic core {\citep[][]{boukare2022deep,boukare2025role}?} Current models of tidal dissipation in rocky exoplanets have mostly presumed the mantle is solid (viscoelastic); if melt is considered, it is only crudely accounted for by adjusting parameters like solid-state viscosity and shear modulus. By contrast, 
\cite{tyler2015tidal}, \cite{hay2020powering}, and
\cite{aygun2024tidal} have shown that Io's hypothesized magma ocean
may not be described as such --- that a fluid treatment may be necessary.

Here we provide such a treatment for short-period exoplanets,
extending our earlier work  \citep{farhat2025tides} to study the time-dependent, fluid-dynamical response of lava worlds to eccentricity tides. We model how a molten magma layer of arbitrary depth flows across the dayside hemisphere, as the planet is flexed repeatedly from apoapse to periapse and back --- we are interested in the dynamical (as opposed to equilibrium) tide, i.e. the ``sloshing'' of the magma ocean  as it is forced by multiple terms in the tidal potential.

Tidally driven variations in surface temperature on sub-orbital and
orbit-to-orbit timescales may be accessible observationally. The ultra-short period planet 55 Cancri e \citep[$P_{\rm orb} = 0.74$ d, $M_{\rm p} = 8.59\pm{0.43} \, M_\oplus$, $R_{\rm p} =
1.947\pm 0.038\, R_\oplus$; ][]{Crida_2018} illustrates what is possible in the time
domain.  Spitzer secondary-eclipse and phase curve measurements
found dayside brightness temperatures spanning 
$1365^{+219}_{-257}$ to $2528^{+224}_{-229}$~K, and a hotspot
that might be displaced eastward of the substellar point by as much as 
$41^\circ\pm 12^\circ$
(\citealt{demory2016variability,demory2016map,tamburo2018confirming};
but see \citealt{mercier2022revisiting} who question the hotspot offset). In the visible, CHEOPS found the phase curve amplitude and phase 
to vary over weeks to months 
\citep[][]{valdes2023investigating}. Most recently, JWST in the near-infrared discovered
rapid, strong, and wavelength-dependent variability in dayside
emission over a few days
\citep[][]{patel2024jwst,hu2024secondary}. Is this remarkable time
variability due to tidal forcing? The published eccentricity $e_{\rm p}=0.05 \pm 0.03$ \citep[][]{bourrier201855} 
is probably consistent with zero;
none of the known planets in the 55~Cnc system can plausibly force a
percent-level eccentricity in planet e \citep{ferraz2025tidal}; and
non-tidal atmospheric dynamics powered by extreme stellar irradiation
offers an alternative route to variability
(\citealt{loftus2025extreme}; see also \citealt{bromley2023chaotic}). Still,
55 Cnc e and other USPs 
\citep[e.g.][]{kreidberg2019absence,crossfield2022gj,greene2023thermal,zhang2024gj,zieba2022k2}
demonstrate the increasing power of time-domain and spectroscopic
studies to characterize planetary surfaces. One day, 
magma oceans and their chemical exchanges with the atmosphere might
come within observational reach \citep[e.g.][]{ito2015theoretical,wordsworth2022atmospheres,zilinskas2023observability,tian2024atmospheric,lichtenberg2024super,Gkouvelis2025}.

This paper is organized as follows. In section \ref{sec:theory}, we
lay out our method for solving the Laplace Tidal Equations governing a dayside magma ocean forced by eccentricity tides
and damped by a simple prescriptive Rayleigh drag. 
 Section \ref{sec:results} presents results, principally how the thermal light curve of the planet and the hotspot longitude can
deviate from predictions of pure stellar insolation
when tides are included. In section \ref{Section_thermal_equilibria}
we relate the surface thermodynamics treated in previous sections to
the thermodynamics of the deep interior. There we show how a fluid
magma ocean may be sustained in thermal equilibrium, and how its depth
can extend to hundreds if not thousands of kilometers. Section \ref{sec:sum}
summarizes and charts future directions.

\section{Theory of Dynamical Lava Tides} \label{sec:theory}
We are interested in modeling the lava flows within a hemispherical magma ocean on the
dayside of a short-period rocky planet, subject to eccentricity  tidal
forcing from the host star. The melt fraction in the silicate ocean is assumed
to be large enough ($\gtrsim 0.2$--0.6 as suggested by experiments;
see e.g.~\citealt{scott2006effect,costa2009model}) that the ocean's
behavior is that of a viscous fluid, even allowing for suspended
solid inclusions \citep[e.g.][]{tonks1993magma,abe1997thermal}.

\subsection{Laplace tidal equations}
We utilize the classical Laplace tidal equations (LTEs) in the strongly viscous, creep flow limit \citep[][]{tyler2015tidal,farhat2025tides}. {On a spherical planet with radius $R_{\rm p}$, harboring a surface magma ocean of uniform thickness $H$,} the linear LTEs comprise a coupled system of momentum and mass conservation \citep[e.g.][]{longuet1968eigenfunctions}:
\begin{subequations}
\label{momentum_continuity}
\begin{align}\label{momentum1} 
    &\partial_t\Vec{u}+\sigma_{\rm R}\Vec{u}+\Vec{f}\cross\Vec{u} = g \grad (\zeta_{\rm eq} - \zeta),\\
    &\partial_t\zeta +\grad\cdot {H\Vec{u}}=0 \label{continuity1}.
\end{align}
\end{subequations}Here $t$ is time; $\Vec{u}= u_\theta\hat{\theta}+u_\lambda \hat{\lambda}$ is the horizontal velocity field of the lava tidal flow in polar coordinates ($\theta$ being the co-latitude and $\lambda$ the longitude);  $g$ is the gravitational acceleration at the surface; $\zeta$ is the tidally induced vertical displacement; $\zeta_\mathrm{eq}$ is the equilibrium tidal displacement; $\grad=R_{\rm p}^{-1}\left[\hat{\theta}\partial_{\theta}+\hat{\lambda}(\sin\theta)^{-1}\partial_{\lambda}\right]$ is the two-dimensional horizontal gradient operator; $f=2\Omega_{\rm p}\cos\theta$ is the Coriolis parameter for a planet rotating with velocity $\Omega_{\rm p}$; and $\sigma_{\rm R}$ is the Rayleigh drag frequency. The latter is the inverse of the timescale over which a tidally excited wave can be brought to rest by 
various damping mechanisms, including friction at the bottom of the oceanic layer,
viscosity, pressure drag against suspended solids, and Darcy-like dissipation when the magma ocean is close to the rheological transition. 

\subsection{Tidal forcing potential}
The tidal forcing from the star enters via the equilibrium tidal displacement $\zeta_\mathrm{eq}$, as given by the equipotential surface induced by the tidal potential $U_{\rm T}$ ($\zeta_\mathrm{eq}=U_{\rm T}/g$). We focus in this work on eccentricity tides. By virtue of the periodic nature of the forcing, the tidal potential is resolved  in time as a Fourier series, 
and in space using Associated Legendre Functions $(P_p^q)$ \citep[e.g.][]{kaula1969introduction}:
\begin{equation}\label{expansion_UT}
    U_{\rm T} = \mathfrak{Re}\left\{\sum_{k=-\infty}^\infty\sum_{p=2}^\infty\sum_{q=0}^n U_{p,q}^{{\rm T}; k} 
    P_p^q(\cos\theta) 
        e^{i[\sigma_q^k t+q\lambda]}\right\},
\end{equation}
where $\sigma_q^k = q\Omega_{\rm p}-k n_{\rm orb}$ is the tidal forcing frequency, $n_{\rm orb}$ is the mean motion of the planet, and 
\begin{equation}\label{Upqs}
   U_{p,q}^{{\rm T}; k}  = \frac{GM_\star}{a_{\rm p}}\left(\frac{R_{\rm p}}{a_{\rm p}}\right)^p A_{p,q,k}(e_{\rm p}).
\end{equation}
Here $G$ is the gravitational constant, $M_\star$ is the mass of the perturbing host star, $a_{\rm p}$ is the planet's semi-major axis, $e_{\rm p}$ is its eccentricity, the dimensionless eccentricity functions 
\begin{align}\nonumber
    A_{p,q,k}(e_{\rm p}) &= (2-\delta_{q,0}\delta_{k,0})(1-\delta_{q,0}\delta_{k<0})\\
    &\times \sqrt{\frac{2(p-q)!}{(2p+1)(p+q)!}}P_p^q(0)X_k^{-(p+1),q}(e_{\rm p}),
\end{align}
where the Kronecker delta function 
$\delta_{i,j}=1$ if $i=j$ and 0 otherwise, and $X_k^{p,q}(e_{\rm p})$ are the Hansen eccentricity functions \citep[e.g.][]{hughes1981computation,laskar2005note}{}{}. We show in Appendix~\ref{Appendix_Hansen_coeffs} our method of computing the Hansen coefficients.

\subsection{Solution strategy}
\citet{farhat2025tides} provided a closed-form analytical solution to the LTEs (Eqs.~\ref{momentum1}-\ref{continuity1}) for a global oceanic shell in the highly damped regime, i.e.~when friction dominates and Coriolis effects are negligible. 
Here, the magma ocean is restricted to a hemispherical shell on the dayside of the planet, and we retain the Coriolis term in the equation. We proceed to solve the LTEs semi-analytically, using an approach similar to that adopted for the Earth's (paleo-)oceans \citep[][]{longuet1968eigenfunctions,webb1980tides,hansen1982secular,tyler2011tidal,farhat2022resonant,auclair2023can}, but with relatively amplified values for $\sigma_{\rm R}$ to 
study the damped, creeping flow regime of the lava fluid. 

With $\vec x$ being the tidally driven horizontal magma displacement (i.e. $\partial_t\vec x=\vec u)$, the momentum equation is rewritten as:
\begin{equation}
 \left[\partial_t^2 +(\sigma_{\rm R}+\Vec{f}\cross)\partial_t \right]\Vec{x}+g \grad(\zeta-\zeta_\mathrm{eq}) =0.
\end{equation}
Following \citet{proudman1920dynamical}, we apply Helmholtz's theorem \citep[e.g.][]{arfken1999mathematical} to decompose  $\vec x$ into  curl-free and divergence-free vector fields:
\begin{equation} \label{horz_tidal_x}
     \Vec{x}=\grad\Phi +\grad\Psi\cross\hat{r},
\end{equation}
where $\Phi$ is a divergent displacement potential function, $\Psi$ is a rotational displacement stream function, and $\hat r$ the radial direction in spherical coordinates.

\subsection{Boundary conditions}
One needs to define the boundary conditions for each component of the flow. {We assume that lava tidal flows are bounded by the day-night terminator. We idealize this boundary as impermeable, in the sense of vanishing normal tidal flow across it. In reality, the terminator is likely to be a finite-width, partially molten transition zone that is mechanically compliant, partially permeable, and dissipative, through phase changes, melt percolation, or mush deformation. The terminator may therefore provide an additional sink of tidal-wave energy and contribute to the effective dissipative coefficient $\sigma_{\rm R}$. For practical purposes here, we ignore such complications and model the terminator as a sharp impermeable interface. }

The boundary condition we impose is  $\vec{x}\cdot\hat{n}=0$, where $\hat{n}$ designates the normal to the terminator lying in the local horizontal plane. {This condition corresponds to an impermeable, free-slip boundary, i.e.~the normal tidal flow vanishes across the terminator, while the tangential flow is not required to do so.} Without another boundary condition allowing us to solve the two fields $\Phi$ and $\Psi$ separately, we follow the tradition of imposing the impermeability condition on both components simultaneously \citep[e.g.][]{gent1983consistent}: $\grad\Phi \cdot\hat{n}=0$ and $\grad\Psi\cross\hat{r} \cdot\hat{n}=0$. The former condition implies that the gradient of $\Phi$ vanishes in the direction normal to the terminator, a Neumann boundary condition; the latter condition  $\grad\Psi\cdot (\hat{r}\cross \hat{n})=0$ implies that the stream function has zero tangential derivative and is therefore constant along the coastline, a Dirichlet boundary condition.

Thus the potential and stream functions, defined on a compact connected domain $\mathcal{O}$ corresponding to the magma ocean basin, satisfy on the ocean coastline $\mathcal{\partial O}$ either a Neumann or Dirichlet boundary condition \citep[][]{unno1979nonradial}. As such, these functions can be expanded in terms of orthogonal eigenfunctions $\{\phi_r,\psi_r\}$
satisfying the wave equations
\begin{subequations}
\label{eigen_equations}
\begin{align}
    &\grad^2\phi_r + \mu_r\phi_r=0,\hspace{1cm}\hat{n}\cdot\grad\phi_r|_{\partial O}=0,\label{wave_equation1}\\
    &\grad^2\psi_r + \nu_r\psi_r=0,\hspace{1cm} \psi_r|_{\partial O}=0\label{wave_equation2},
\end{align}
\end{subequations}with the horizontal Laplacian defined as $\grad^2=(R_{\rm p}\sin\theta)^{-2}\left[\sin\theta\partial_{\theta}\left(\sin\theta\partial_\theta\right)+\partial_{\lambda}^2\right]$. Here the subscript $r$ is a harmonic index (more on this label below). With $\Psi$ having zero tangential derivative on the terminator,  we have set the Dirichlet boundary value to $\psi_r = 0$. 

\subsection{Series solutions}
\begin{figure*}[t]
\includegraphics[width=\textwidth]{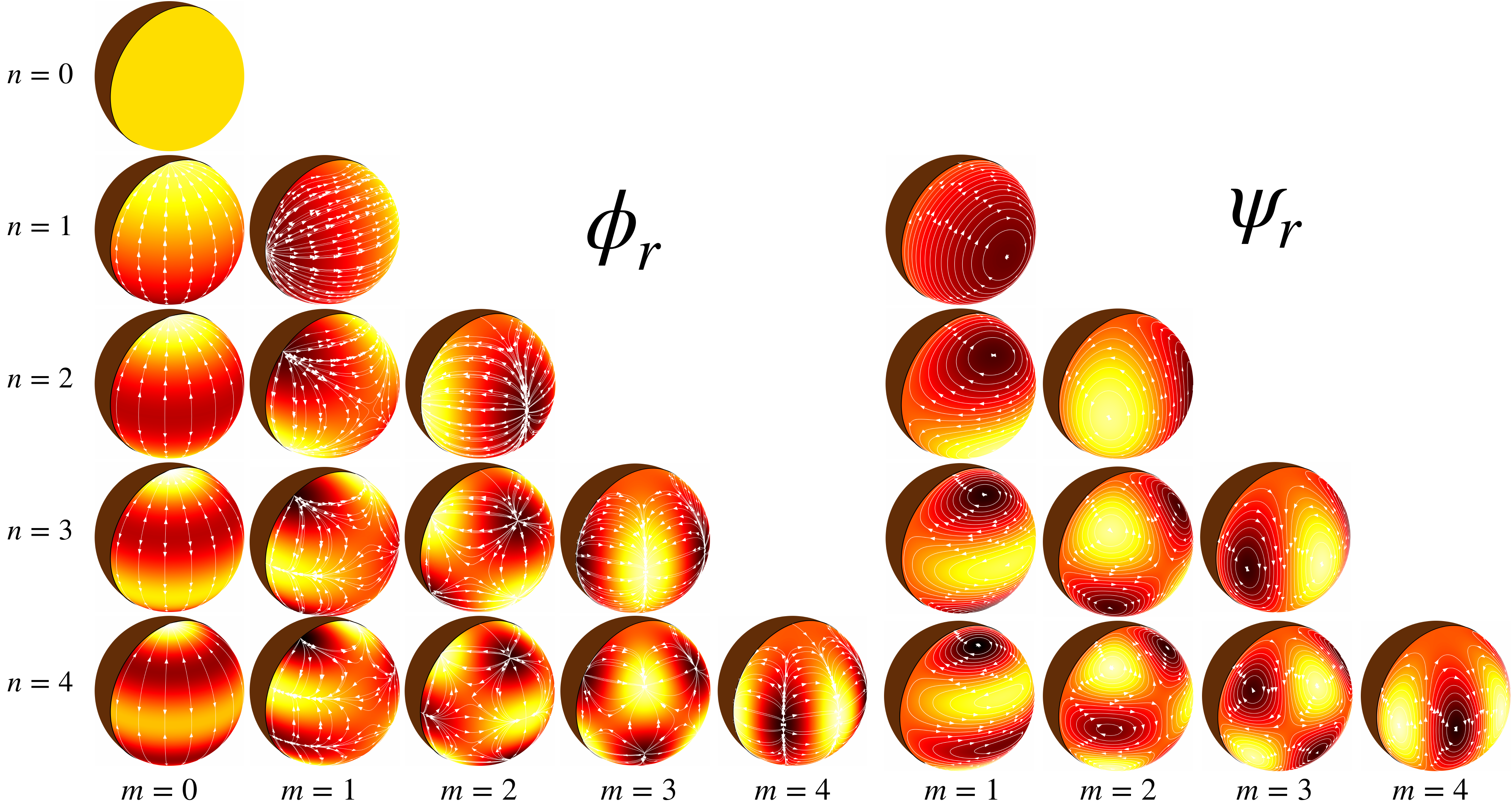}\caption{Eigenfunctions describing tidal lava flows on a hemispherical dayside magma ocean. The nightside is masked in brown. Left half shows the scalar potential functions $\phi_{r}$, computed by taking the real part of \eq{phi_r_eq} and plotted for $n=0,...,4$ and $m=0,...,n$. Right half shows the stream functions $\psi_r$ computed by taking the real part of \eq{psi_r_eq} and plotted for $n=1,...,4$ and $m=1,...,n$. Bright and dark colors correspond to negative and positive values of the eigenfunctions, respectively. Left streamlines follow the tidal irrotational flow $\grad\phi_r$, and right streamlines the tidal non-divergent flow $\grad\psi_r\cross\hat{r}$.}
\label{Fig_Phi_Psi}
\end{figure*}

The solutions to the latitudinal components of Eqs.~(\ref{eigen_equations}) are the Associated Legendre Functions $P_n^m$, while the longitudinal solutions take the form 
$\exp( im\lambda)$:
\begin{subequations} \label{phi_psi}
\begin{align}\label{phi_r_eq}
    &\phi_r = \frac{\alpha_{nm}}{R_{\rm p}}P_{n}^m(\cos\theta)\exp( im\lambda),\\\label{psi_r_eq}
    &\psi_r = -i \frac{\alpha_{nm}}{R_{\rm p}}P_{n}^m(\cos\theta)\exp( im\lambda).
\end{align}
\end{subequations}
We have chosen $(n,m)$ to expand the harmonics of the tidal response, as distinct from the $(p,q)$ labels used earlier to expand the forcing (Eq.~\ref{expansion_UT}).  Each $(p,q)$ forcing harmonic excites a mix of different $(n,m)$ response harmonics.\footnote{In the case of a spherical fluid shell, the response is expanded in a series of Hough modes \citep{hough}, where for a given tidal forcing mode $(p,q)$ the shell responds as a sum of modes with the same $m=q$ but various $n$-modes \citep[e.g.][]{auclair2018oceanic,farhat2022resonant}. The one-to-one forcing-to-response modal mapping characteristic of pure solid tides is broken by the Coriolis force acting on the fluid. For our hemispherical shell, the zonal symmetry on the sphere is also broken, so each forcing mode excites a spectrum of basin modes with different $n$'s and $m$'s. Fortunately, the coastline living on a great circle preserves the separability of the solution, allowing us to maintain the Legendre polynomials as basis expansion functions. This separability is lost when one considers arbitrary coastlines \citep[]{auclair2023can} or misalignments in the system's spin-orbit geometry \citep[][]{auclair2025anisotropic}. \label{footnote_modes}}
Inserting (\ref{phi_psi}) into (\ref{eigen_equations}) yields the eigenvalues $\mu_{r}=\nu_r=n(n+1)/R_{\rm p}^2$ and the normalization coefficient
\begin{equation}
\alpha_{n,m}=\sqrt{\frac{(2n+1)}{\pi}\frac{(n-m)!}{(n+m)!}\frac{1}{1+\delta_{m,0}}}.
\end{equation}
Every subscript $r$ denotes 
a unique pair of $n$ and $m$. For $\phi_r$, $n =\{0,1,2,\dots\}$ and $m=\{0,1,...,n\}$, while for $\psi_r$, $n=\{1,2,3,\dots\}$ and $m=\{1,2,...,n\}$. We follow the scheme detailed in \citet[their section 4]{longuet1970free} to map $r$'s to pairs $(n,m)$. It is possible to split the functions $(\phi_r,\psi_r)$ into two sets describing tidal flows that are either symmetric or anti-symmetric about the equator. In Figure~\ref{Fig_Phi_Psi} we plot these hemispherical basis functions and we show the streamlines of the corresponding flows. 

Introducing the time-dependent coefficients $p_r(t)$ and $p_{-r}(t)$, we write the potential and stream functions as:
\begin{subequations}
\label{functions_expansion}
\begin{align}
&\Phi(\theta,\lambda,t)=\sum_{r=1}^{\infty}p_r(t)\phi_r(\theta,\lambda),\\
   &\Psi(\theta,\lambda,t)=\sum_{r=1}^{\infty}p_{-r}(t)\psi_r(\theta,\lambda).
\end{align}
\end{subequations}Note that despite the (historical) notation which might suggest otherwise, $p_r$ and $p_{-r}$ are independent coefficients, i.e., in general, one cannot be derived from the other. Substituting equations \eqref{eigen_equations} and \eqref{functions_expansion} into the continuity equation \eqref{continuity1}, we recover the vertical tidal displacement of the lava waves:
\begin{equation}\label{eq.zetasol}
  \zeta=H\sum_{r=1}^\infty \mu_r p_r\phi_r ,
\end{equation}
for constant free parameter $H$. 
Finding the tidal flow patterns reduces to solving for the time-dependent coefficients $p_r$ and $p_{-r}$. We delegate the details of the solution method to Appendix~\ref{LTE_Solution}; the executive summary is that we  
substitute all the aforementioned expansions into the LTEs (Eqs.~\ref{momentum_continuity}), 
take the dot product of the momentum equation with $\grad\phi_r$ to obtain one set of equations, repeat with $\grad\psi_r\cross\hat{r}$ to obtain a second set, and integrate over the oceanic domain  $\mathcal{O}$ to obtain a coupled linear system for the coefficients $p_r$ and $p_{-r}$: 
\begin{align}\nonumber
        -\sigma^2p_r-i\sigma\sigma_{\rm R} p_r +gH\mu_r p_r 
        -g{\zeta}_{\rm eq;r} \\
        - \frac{2i\sigma\Omega_{\rm p}}{\mu_r}\sum_{s=-\infty}^{\infty}\beta_{r,s}p_s  & =0,\label{lin_sys_1}\\
        -\sigma^2p_{-r} -i\sigma\sigma_{\rm R}p_{-r} -\frac{2i\sigma\Omega_{\rm p}}{\nu_r}\sum_{s=-\infty}^{\infty}\beta_{-r,s}\partial_t p_s & =0.\label{lin_sys_2}
\end{align}
We have used $\partial_t\rightarrow i\sigma_q^k$ (Eq.~\ref{expansion_UT}) and replaced $\sigma_q^k$ with $\sigma$ for short. The above system is solved for all $p_r$ and $p_{-r}$ for a given $\sigma$. The variable $\zeta_{\rm eq;r}$ corresponds to the $r$th harmonic of the equipotential surface, 
i.e.:
\begin{equation}
    \zeta_{\rm eq;r}= \int_{\mathcal{O}} \phi_r \zeta_{\rm eq} dA,
\end{equation}
with $dA$ denoting an area element of the ocean domain.
The gyroscopic coefficients $\beta_{r,s}$ (Eqs. \ref{gyro1}) encode the effect of rotational distortion to the tidal waves and the effect of the boundary conditions imposed by the terminator. We ignore here the effect of solid deformation on the magma ocean dynamics, which can enter through the effects of self-attraction and loading between the ocean and the underlying deforming solid mantle \citep[e.g.][]{unno1979nonradial}. These effects are secondary and can be absorbed into the free parameter $\sigma_{\rm R}$ \citep[see e.g.][]{farhat2022resonant,auclair2023can}.

\subsection{Summary of solution method}
With our problem formally set up, 
we solve for the tidal lava flow as follows:
\begin{itemize}
    \item[(\textit{i})]A 2-D mesh grid representing the dayside surface of the planet is established.
    \item[(\textit{ii})] On each grid point, and for a given tidal forcing frequency 
    $\sigma$, the equilibrium tide is defined, and the system of Equations (\ref{lin_sys_1})--(\ref{lin_sys_2}) is solved. In practice, the infinite system must be truncated 
    at a certain order $r_{\rm max}$. We test the solution dependence on $r_{\rm max}$ and find that $r_{\rm max}=100$ suffices.
     \item[(\textit{iii})] With the coefficients $p_r$ and $p_{-r}$ solved for at a specific forcing frequency, we recover the potential and stream functions $\Phi$ and $\Psi$, respectively.
    \item[(\textit{iv})] Taking the gradients of these functions, we recover the complex displacement vector field 
    $\hat{x}(\sigma,\theta,\lambda)=\grad\Phi +\grad\Psi\times\hat{r}$. We use the hat symbol on $\hat{x}$ to denote that this is just one component of the displacement field specific to $\sigma$; later in step (vi) we will sum over all $\sigma$.
    \item[(\textit{v})] The complex velocity vector field is obtained by taking 
    $\hat{u}(\sigma,\theta,\lambda) = i\sigma
    \hat{x}(\sigma,\theta,\lambda)$. 
     \item[(\textit{vi})]The
     time-dependent velocity field is finally computed by summing over the Fourier spectrum and taking the real part:
     \begin{equation}
         \vec{u}(\theta,\lambda,t)=\mathfrak{Re}\left\{ \sum_{k=-\infty}^\infty \sum_{q=0}^p\hat{u}(\sigma_q^k,\theta,\lambda)\exp(i\sigma_q^kt)\right\}.
     \end{equation}
\end{itemize} 

\subsection{Evaluating the tidally dissipated power}
With the velocity field in hand, the thin-shell Rayleigh drag produces a stress (force per area) of $\boldsymbol{\tau}_{\rm R}=-\rho_{\rm m} H\,\sigma_{\rm R}\,\vec{u}$, where $\rho_{\rm m}$ is the magma ocean's density. The instantaneous rate at which this stress does work on the flow, i.e.~the tidal dissipation per unit area, is 
\begin{align}\nonumber
    \mathcal{P}_{\rm T}(\theta,\lambda,t) &= \boldsymbol{\tau}_{\rm R}\cdot \vec{u}\\
    &=\rho_{\rm m}H\sigma_{\rm R}\vec{u}(\theta,\lambda,t)
    \cdot\vec{u}(\theta,\lambda,t), \label{inst_PT}
\end{align}
The total tidal heating in the ocean basin is obtained by integrating over the domain $\mathcal{O}$, and averaging over
a particular interval of time:
\begin{equation}\label{average_PT}
    \overline{\mathcal P}_{\rm T} = \left\langle\int_{\mathcal{O}}\rho_{\rm m}H\sigma_{\rm R}\vec{u}(\theta,\lambda,t)\cdot\vec{u}(\theta,\lambda,t) dA\right\rangle.
\end{equation}
In practice we take the time average $\langle ... \rangle$ over 10 orbital periods.

In addition to tidal heating, we also compute the stellar insolation,  modeling the dayside as a gray Lambertian surface. As the planet orbits its star, at each time $t$, we solve Kepler’s equation to obtain the eccentric anomaly $E_{\rm p}$ and the star-planet distance
\begin{equation}
 r_{\rm p}(t) = a_{\rm p}\,(1 - e_{\rm p} \cos E_{\rm p}).   
\end{equation}
The bolometric stellar flux is then
\begin{equation}
F_{\star}(t) = \frac{L_\star}{4\pi\,r_{\rm p}^2(t)} 
= \sigma_{\rm SB}\,T_\star^4 \left(\frac{R_\star}{r_{\rm p}(t)}\right)^2 ,
\end{equation}
where $\sigma_{\rm SB}$ is the Stefan-Boltzmann constant, and $L_\star$, $R_\star$, and $T_\star$ are the luminosity, radius, and effective temperature of the host star. The cosine of the local stellar zenith angle is
$\mu_0(\theta,\lambda)=\sin\theta\,\cos\lambda$ (where the substellar point is located at $\theta=\pi/2$ and $\lambda=0$). The local absorbed insolation flux is then
\begin{equation}
\mathcal{P}_{\rm ins}(\theta,\lambda,t) 
= (1-A_{\rm B})\,F_{\star}(t)\,\mu_0(\theta,\lambda)\,\mathcal{H}\!\left[\mu_0(\theta,\lambda)\right],
\end{equation}
where $A_{\rm B} = 0.1$ is the Bond albedo \citep[][]{essack2020low} and $\mathcal{H}$ is the Heaviside step function that sets the nightside ($\mu_0\le 0$) to zero incident flux. We have neglected any dependence on wavelength, reflectance angle, stellar limb darkening, and atmospheric scattering. 
The effective surface temperature $T_{\rm s}$ assumes blackbody radiation and full redistribution on the planet: 
\begin{equation}
    T_{\rm s} 
\equiv 
    \left(\frac{\overline{\mathcal{P}}_{\rm T}+\overline{\mathcal{P}}_{\rm ins}}{4\pi R_{\rm p}^2\epsilon\sigma_{\rm SB}} \right)^{1/4} \label{eq:Tsurfacerad}
\end{equation}
where $\epsilon$ is the planetary emissivity \citep[${\sim}0.9$ in the infrared; e.g.][]{henderson1996new}{}{}.

\section{Lava Tidal Waves}\label{sec:results}
We explore the effect of tidal lava flows on the time-dependent thermal evolution of the planet surface.  We first examine the global time-and-space averaged tidal response as a function of the free parameters.

\subsection{Global time-averaged tidal response}
We use planet parameters inspired by 55~Cancri~e \citep[e.g.][]{winn2011super,Crida_2018}  
to define our fiducial system. Unless stated otherwise,
we set $M_{\rm p}=8.6 M_\oplus$, 
$R_{\rm p} = 1.9 R_\oplus$, 
$P_{\rm orb} = 0.74$ d, 
$a_{\rm p}=0.016$~AU, 
$M_\star=1 M_\odot$,
$R_\star=1 R_\odot$,
$T_\star=5200$~K, 
and $L_\star=4\pi R_\star^2\sigma_{\rm SB}T_\star^4$.
Our main free parameters are: (\textit{i}) the orbital eccentricity $e_{\rm p}$ 
which controls the strength of the tidal forcing;  (\textit{ii}) the frequency $\sigma_{\rm R}$ characterizing the efficiency of wave damping; and (\textit{iii}) the magma ocean's thickness $H$.

\begin{figure}[t]
\includegraphics[width=.5\textwidth]{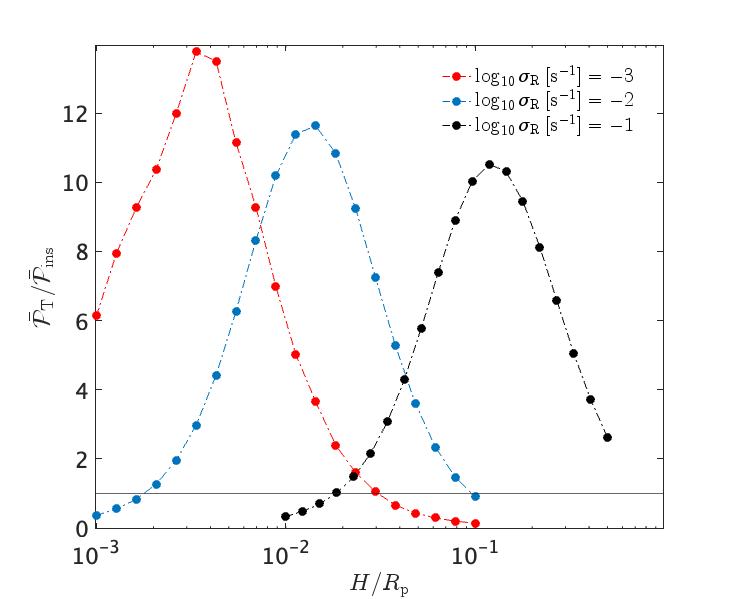}
\caption{{Time-and-space averaged heat generated by tides in the magma ocean of our fiducial short-period planet (modeled after 55 Cancri e) compared to the average stellar insolation, assuming an orbital eccentricity of $e_{\rm p}=0.05$ and three values for the Rayleigh drag frequency $\sigma_{\rm R}$. The tidal power peaks and exceeds the stellar insolation for an ocean thickness $H_{{\rm max}\,\overline{\mathcal{P}}_{\rm T}} $, defined by \eq{H_peak}.} }
\label{Fig_PT_over_Pins}
\end{figure}

In Figure~\ref{Fig_PT_over_Pins} we show, at fixed $e_{\rm p} = 0.05$, how much more power is produced by tides ($\overline{\mathcal{P}}_{\rm T}$) relative to stellar insolation  ($\overline{\mathcal{P}}_{\rm ins}$) as a function of $H$, for three values of $\sigma_{\rm R}$. Whether any of our choices for $\sigma_{\rm R}$ is realistic is not completely clear. As we are modeling creep flow in magma oceans, our chosen values for $\sigma_{\rm R}$ are larger than those used for modeling water oceans \citep[e.g.][]{garrett1971age,webb1973age,auclair2023can}, which seems appropriate as lava waves should damp faster than water waves. In the limit of larger  $\sigma_{\rm R}$, one loses the dynamical tide and recovers the equilibrium tide of a rigid shell with a Darwin lag angle between the tidal response and the forcing potential.

The tidal heating is seen in Fig.~\ref{Fig_PT_over_Pins} to peak at an intermediate $H$, and to drop below stellar heating if either
$H\rightarrow0$ or $H\rightarrow R_{\rm p}$. The ocean thickness appears critical to determining whether the planet's surface energetics are dominated by tides or the stellar flux.

To understand why $\overline{\mathcal{P}}_{\rm T}$ peaks, we simplify the LTEs, keeping only a few ingredients: linear momentum with Rayleigh drag, hydrostatic pressure, and a forcing term $\hat{f}_{\rm T}$ having only one frequency $\sigma$; plus mass continuity, all for just one horizontal mode of spherical degree $\ell$. The momentum equation simplifies to:
\begin{equation}
    \left(-i\sigma + \sigma_{\rm R}\right)\hat{u} = -i gk_\ell\hat\zeta + \hat{f}_{\rm T},
\end{equation}
where we used $\partial_t\rightarrow i\sigma$ and $\grad \rightarrow ik_{\ell}$, with $k_\ell$ being the spatial wavenumber of the forcing mode  ($k_\ell=\sqrt{\ell(\ell+1)}/R_{\rm p}$). The continuity equation becomes
\begin{equation}
    -i\sigma\hat\zeta + iHk_\ell \hat{u}=0.
\end{equation}
The above two equations combine to yield the velocity:
\begin{equation} \label{eq:u_hat_simple}
    \hat{u} = \frac{\sigma \hat{f}_{\rm T}}{gHk_\ell ^2 -\sigma^2+i\sigma\sigma_{\rm R}},
\end{equation}
which is similar in form to the displacement amplitude (but not the velocity amplitude) of a damped forced oscillator with natural frequency $c_{\rm g}k_\ell$, where $c_{\rm g}\equiv\sqrt{gH}$ is the phase speed of surface gravity waves in the shallow water limit. 
The time-averaged dissipated power is
\begin{equation}
 \overline{\mathcal{P}}_{\rm T}=\rho_{\rm m}H \sigma_{\rm R}|\hat{u}|^2 = \rho_{\rm m}H\sigma^2\sigma_{\rm R}\frac{ |\hat{f}_{\rm T}|^2}{(gHk_\ell ^2 -\sigma^2)^2+(\sigma\sigma_{\rm R})^2} 
\end{equation}
from which we can discern various limiting behaviors: 
\begin{itemize}
    \item For $H\rightarrow0$, the ocean has negligible mass, and the prefactor $\rho_{\rm m} H$ causes $\overline{\mathcal{P}}_{\rm T}\propto H\rightarrow 0$. 

    \item For sufficiently large $H$, the natural mode frequency $c_{\rm g}k_\ell$ becomes larger than the damping frequency $\sigma_{\rm R}$ (which in turn is larger than the forcing frequency $\sigma$, by construction).  In the limit that the natural frequency dominates the denominator of (\ref{eq:u_hat_simple}), $|\hat{u}|\propto  H^{-1}$ and $\overline{\mathcal{P}}_{\rm T}\propto H(1/H^2) \propto H^{-1}$,  which explains why the curves decline toward larger $H$ in Figure \ref{Fig_PT_over_Pins}. In this regime,  waves traveling at speed $c_{\rm g}$ cross a wavelength $1/k_\ell$ in a time $1/(k_\ell c_{\rm g})$ that is much shorter than either the forcing period $1/\sigma$ or the damping timescale $1/\sigma_{\rm R}$.

    \item For $H$ between these limits, the dissipated power reaches an extremum at  
    \begin{equation}\label{H_peak}
        H_{{\rm max}\,\overline{\mathcal{P}}_{\rm T}} =  \frac{\sigma^2}{gk_{\ell}^2}\sqrt{1+\left(\frac{\sigma_{\rm R}}{\sigma}\right)^2}.
    \end{equation}
    Many studies (e.g.~\citealt{tyler2011tidal,kamata2015tidal,auclair2019final}) are concerned with
    the weakly damped limit ($\sigma_{\rm R}\ll\sigma)$ where \eq{H_peak} implies the resonance condition $\sigma \approx \sqrt{gHk_\ell^2}$, when the forcing frequency matches the shallow water wave frequency.
    By contrast, we are here concerned with the limit of strong damping, $\sigma_{\rm R}\gg \sigma$, whence \eq{H_peak} implies 
\begin{equation}\label{eq_peak_frequency}
        \left.\sigma\right|_{{\rm max}\,\overline{\mathcal{P}}_{\rm T}} \approx \frac{(c_{\rm g}k_\ell)^2}{\sigma_{\rm R}}, 
    \end{equation}
     and $H_{{\rm max}\,\overline{\mathcal{P}}_{\rm T}} \propto \sigma_{\rm R}$ at fixed $\sigma$, roughly consistent with Figure \ref{Fig_PT_over_Pins}. 
\end{itemize}

For an analogy describing these limits, one can imagine shaking a pan of honey at fixed   cadence. When the layer is very thin, it moves with the pan like a rigid sheet; there is insufficient depth to build surface slope or shear. When the layer of honey is very deep, most of it idles while a thin viscous layer at the bottom of the pan is stirred.  
At some intermediate thickness, the whole column moves and shears coherently, maximizing the work done per cycle and thus the dissipated power.

\begin{figure*}[ht]
\includegraphics[width=\textwidth]{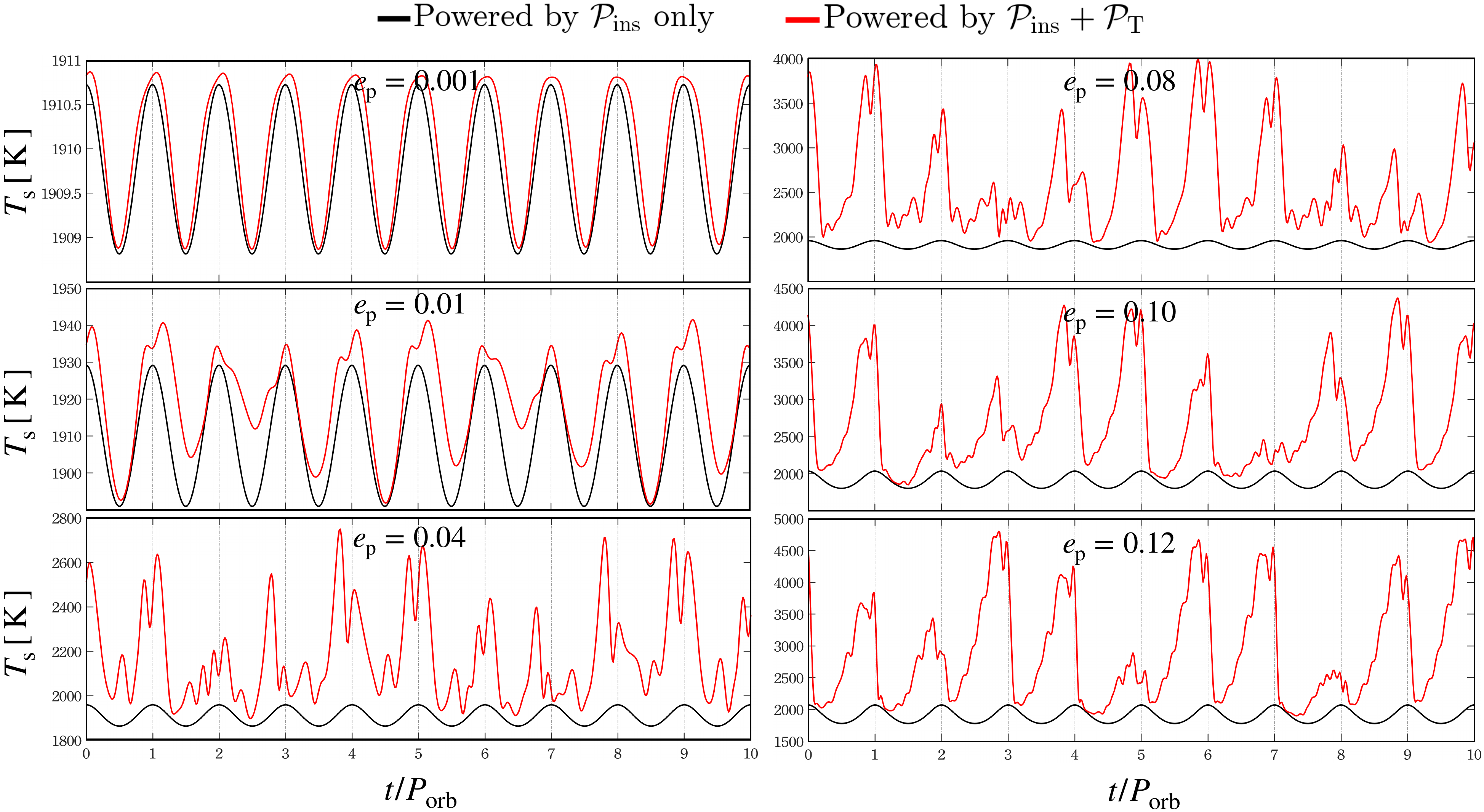}\caption{The effect of lava tidal flows on the thermal emission of our fiducial short-period rocky planet. The surface equilibrium temperature $T_{\rm s}$ is plotted vs.~time accounting for either stellar insolation alone (black curve), or insolation plus tidal heating within the magma ocean (red curve). Orbital eccentricity $e_{\rm p}$ varies between different panels (note different y-axis scales). These results presume $H/R_{\rm p} = 1\%$ and $\sigma_{\rm R} = 10^{-3}~{\rm s}^{-1}$, which according to Fig.~\ref{Fig_PT_over_Pins} implies $\overline{\mathcal{P}}_{\rm T}/\overline{\mathcal{P}}_{\rm ins} \simeq 5$ (tidal heating dominates) for $e_{\rm p} = 0.05$. Whether $\overline{\mathcal{P}}_{\rm T}/\overline{\mathcal{P}}_{\rm ins} > 1$ or $<1$ in thermal equilibrium depends on $e_{\rm p}$ and other model details (see section \ref{Section_thermal_equilibria} for the equilibrium theory).
}
\label{Fig_Ts_variations}
\end{figure*}

\subsection{Thermal variability on orbital timescales}\label{subsec:var}
We now examine the temporal evolution of the surface temperature as a result of dayside lava flows. In Figure~\ref{Fig_Ts_variations}, we set $H/R_{\rm p}=1\%$ (corresponding to a surface magma ocean of depth ${\sim}100$~km) and  
$\sigma_{\rm R}=10^{-3}$~s$^{-1}$ (placing the system in the overdamped regime).  For these parameter choices, independent of $e_{\rm p}$, the right-hand-side frequency of \eq{eq_peak_frequency} is $(c_{\rm g}k_2)^2/\sigma_{\rm R} \simeq 10^{-4}~{\rm s}^{-1}$ for $k_\ell = k_2 = \sqrt{6}/R_{\rm p}$. This frequency is close to the orbital frequency $n_{\rm orb} \simeq 10^{-4}~{\rm s}^{-1}$. Thus the condition of peak dissipation in \eq{eq_peak_frequency} should be satisfied by eccentricity tidal forcing terms having $|q-k|$ of order unity, as the forcing frequency is $\sigma_q^k=q\Omega_{\rm p}-k\,n_{\rm orb}$, and $\Omega_{\rm p}$ is close to $n_{\rm orb}$ (near synchronicity). Thus we expect forcing terms like 
$(q,k)=(2,3)$ and $(2,1)$, and to a lesser degree $(3,1)$ and $(2,4)$, to play a role. Whether a given forcing term matters will depend principally on the strength of its Hansen coefficient, which depends on eccentricity $e_{\rm p}$.

The six panels in Figure \ref{Fig_Ts_variations} vary $e_{\rm p}$. The black curves plot the equilibrium surface temperature obtained by taking into account only stellar insolation. The red curves add lava tidal heating to stellar insolation. We present the evolution over ten orbital periods, initializing the planet at periastron passage. The light curve behaviors are summarized as follows:

\begin{itemize}
\item {\textit{Low eccentricity:}} For $e_{\rm p}= 10^{-3}$, tidal heating enhances the surface temperature from stellar insolation by just a fraction of a degree. In this regime, the eccentricity-dependent Hansen coefficients for $q\neq k$ are too small for the corresponding tidal forcing terms to matter. The Hansen coefficient for the semidiurnal tide $(q,k)=(2,2)$ is large and does not depend on $e_{\rm p}$, but because the planet is spinning nearly synchronously with its orbit, the forcing frequency of $(2,2)$ is much too low to satisfy the peak dissipation condition \eq{eq_peak_frequency}. The resulting tidal horizontal flow is modest, with ${u}_{\rm rms}\lesssim0.1~{\rm m\,s^{-1}}$, comparable to the velocity at which lava flows advance after typical Hawaiian eruptions \citep[‘a‘ā flows on gentle slopes, or pāhoehoe breakouts on steeper slopes; e.g.][]{kauahikaua2003hawaiian}. 

\begin{figure}[t]
\includegraphics[width=.47\textwidth]{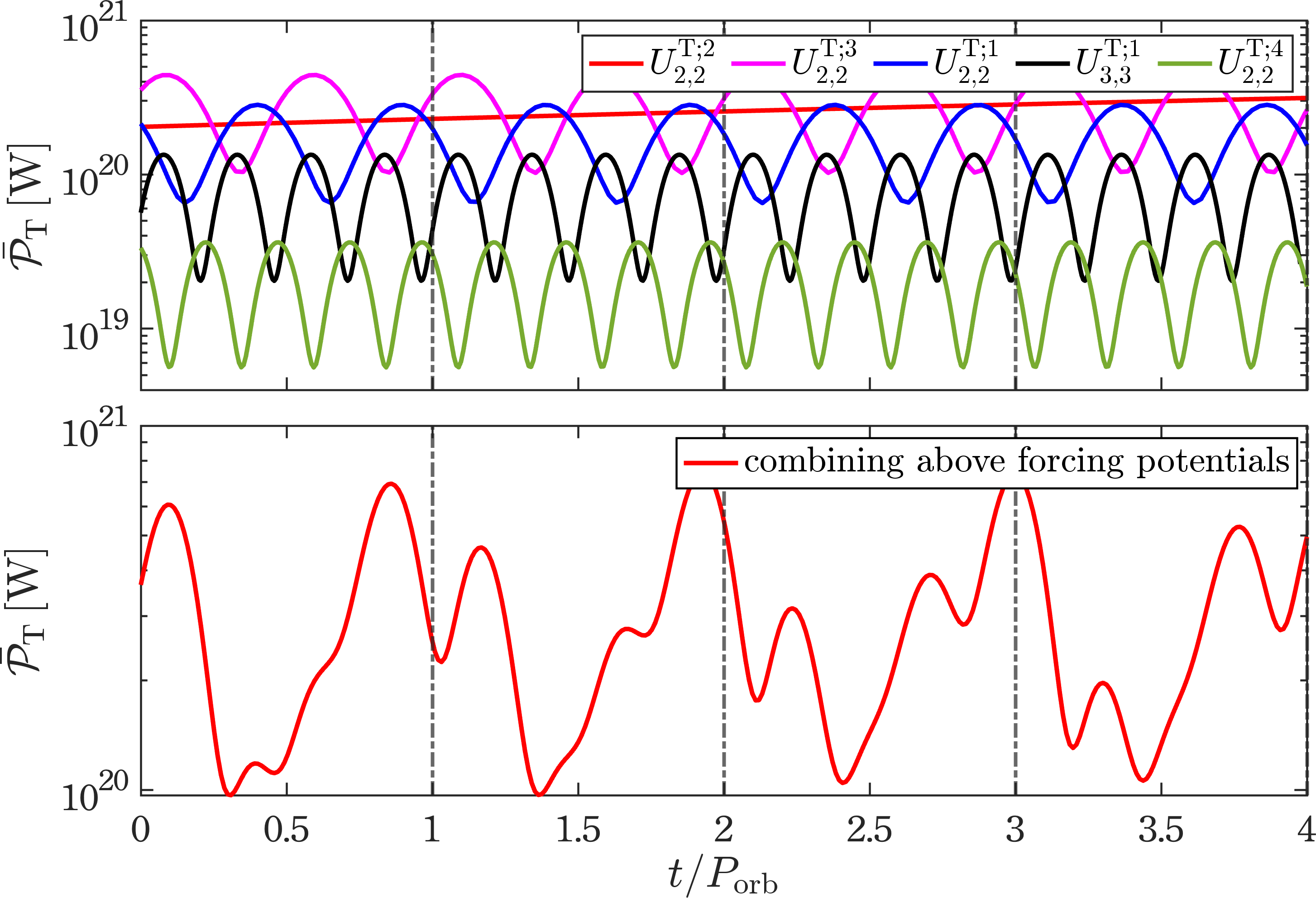}
\caption{Modal analysis of the tidally dissipated power, for $e_{\rm p} = 0.05$, showing the dissipation when individual forcing potentials act in isolation (upper panel) and when they combine (lower panel). A single forcing potential of period $P$ yields strictly periodic dissipation with period $P/2$, because dissipation is quadratic in flow velocity (except for the $p=q=k=2$ component, which drives a nearly steady flow in the rotating frame). Lava velocities $\vec{u}$ driven by different modes superpose linearly, whereas the dissipated power $\propto |\vec{u}|^2$ contains cross terms between modes which beat against each other; the resultant power is no longer strictly periodic.}
\label{indiv_modes}
\end{figure}

\begin{figure}[t]
\includegraphics[width=.47\textwidth]{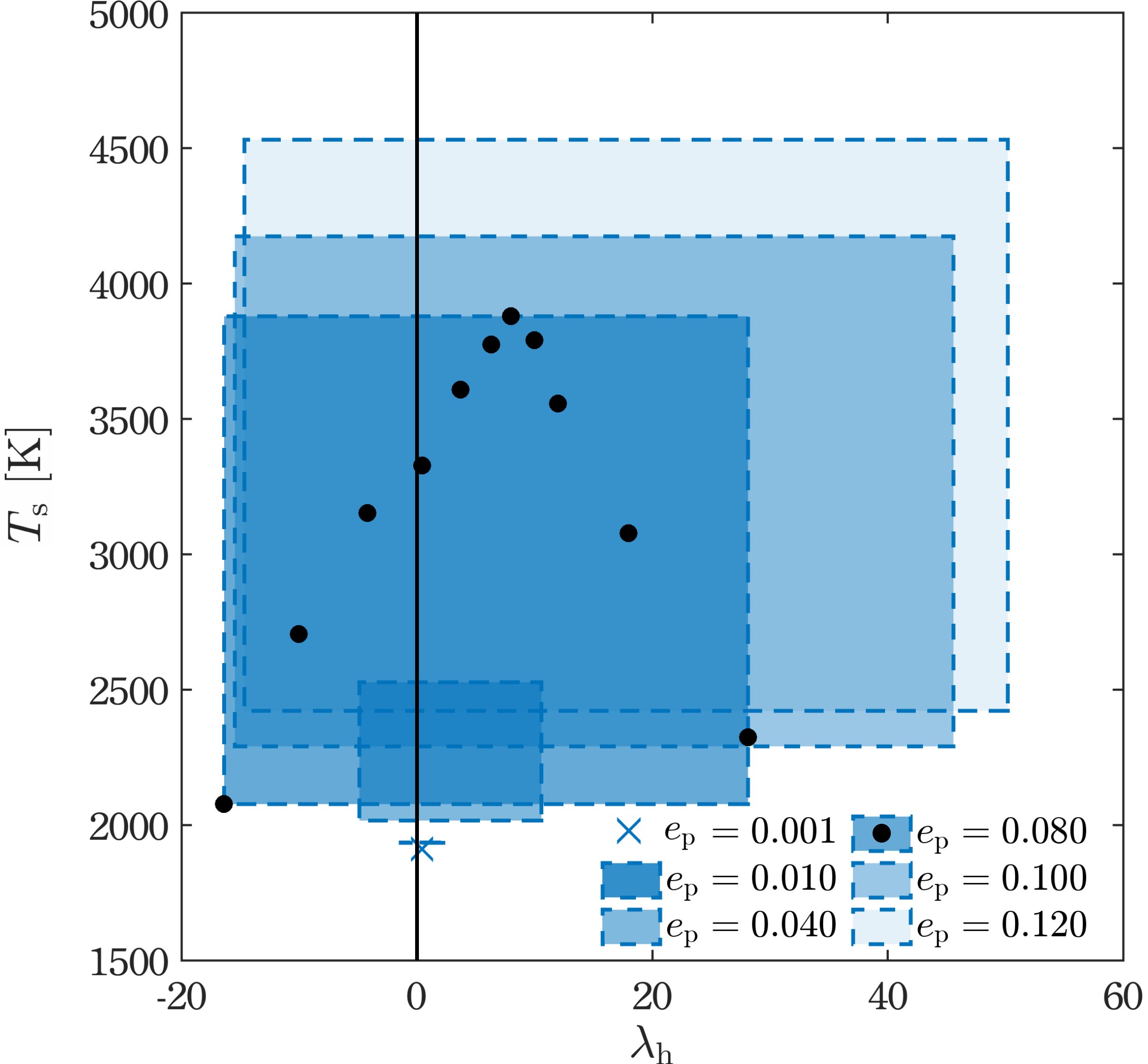}
\caption{Range of variations in surface temperature $T_{\rm s}$ and hotspot longitude $\lambda_{\rm h}$ caused by lava tidal flows.   Different rectangles correspond to different orbital eccentricities $e_{\rm p}$ and show the ranges over which $T_{\rm s}$ and $\lambda_{\rm s}$ vary for our fiducial super-Earth, as sampled stroboscopically at multiple integers of the orbital period $P_{\rm orb}$ over $10P_{\rm orb}$ (see vertical lines in Fig.~\ref{Fig_Ts_variations}). Actual sampled values for $e_{\rm p} = 0.08$ are shown as points. The hotspot is located mostly east, but occasionally west, of the substellar longitude. At $e_{\rm p}=10^{-3}$ heating is dominated by insolation so that thermal variability is negligible, and the rectangle shrinks to a point at $(\lambda_{\rm h},T_{\rm s})\simeq (0^\circ, 1910~{\rm K})$.  }
\label{Fig_hotspot}
\end{figure}

\item \textit{Higher eccentricity}: Tidal heating increases with increasing $e_{\rm p}$. The forcing frequency 
$\sigma_2^2=2\Omega_{\rm p}-2\,n_{\rm orb}$ of the semidiurnal tide 
increases as the pseudo-synchronous spin frequency \citep[e.g.][]{hut1981tidal} 
\begin{equation}\label{Eq_pseudosynchronous}
\Omega_{\rm p}(e_{\rm p}) = n_{\rm orb} \,\frac{
1 + {15}/{2}\,e_{\rm p}^{2} + {45}/{8}\,e_{\rm p}^{4} + {5}/{16}\,e_{\rm p}^{6}}
{\Bigl( 1 + 3\,e_{\rm p}^{2} + {3}{8}\,e_{\rm p}^{4} \Bigr)\,\left(1 - e_{\rm p}^{2}\right)^{3/2}
}
\end{equation}
increases away from $n_{\rm orb}$ with increasing $e_{\rm p}$. Thus semidiurnal forcing comes closer to satisfying the peak dissipation condition (\ref{eq_peak_frequency}); although for $e_{\rm p}$ of a few percent, $\sigma^2_2$ is still a couple orders of magnitude short of $(c_{\rm g}k_2)^2/\sigma_{\rm R}$, the large  $e_{\rm p}$-independent Hansen coefficient of $(2,2)$ renders the semidiurnal tide energetically significant. Other $k$-terms also have non-negligible Hansen weights; there are multiple forcing modes at play, each mode driving a spectrum of tidal waves produced by scattering off the terminator (see footnote \ref{footnote_modes}). When lava waves of different frequencies constructively interfere, there are intense heat pulses (the dissipated power is quadratic in the velocity). Flow velocities increase to ${u}_{\rm rms} \sim 2~{\rm m\,s^{-1}}$ for $e_{\rm p}=0.04$, and ${u}_{\rm rms} \sim 6~{\rm m\,s^{-1}}$ for $e_{\rm p}=0.08$, comparable to basaltic lava speeds in well-developed channels during effusive eruptions on Earth \citep[e.g.][]{kauahikaua2003hawaiian}. Conversely, when the waves destructively interfere, 
heating can return to being dominated by stellar insolation. Thermal contrasts can be dramatic: dayside surface temperatures can vary by of order 1000~K within less than an orbital period for $e_{\rm p}$ as small as $6\%$. The large variations depend on tidal heating dominating stellar insolation ($\overline{\mathcal{P}}_{\rm T}/\overline{\mathcal{P}}_{\rm ins} > 1$), a condition that depends not only on $e_{\rm p}$ but also $H$, $\sigma_{\rm R}$, and other model details.

\begin{figure*}[t]
\includegraphics[width=\textwidth]{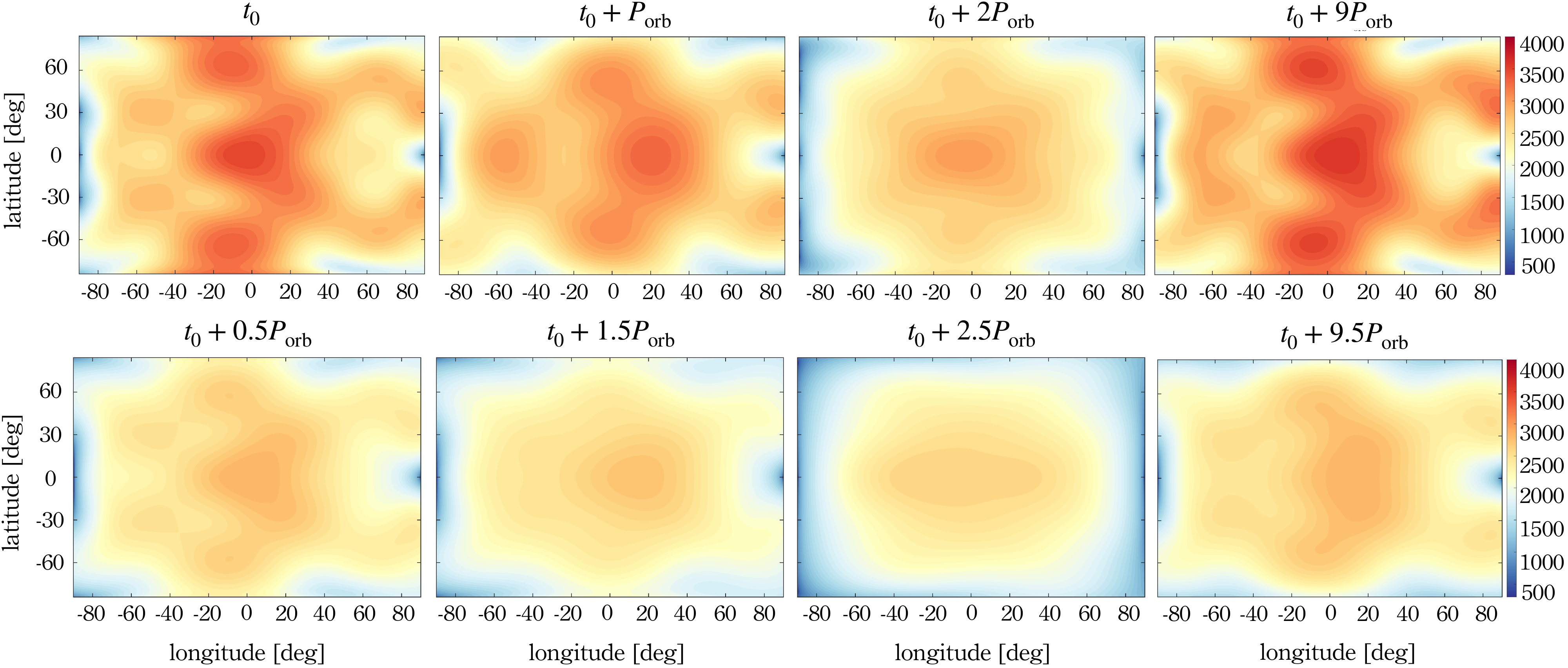}\caption{Snapshots of the surface temperature $T_{\rm s}(\theta,\lambda,t)$~[K] across the dayside of our fiducial planet, captured at multiple pericenter passages (upper row) and apocenter passages (bottom row). We adopt the same parameters used in Figure~\ref{Fig_Ts_variations}, with $e_{\rm p}=0.04$.} 
\label{Fig_snapshots_Ts}
\end{figure*}

\item \textit{Aperiodicity:} 
Interference between lava tidal waves introduces another feature: the
loss of strict periodicity. We analyze this behavior in Figure
\ref{indiv_modes}, for an assumed $e_{\rm p} = 0.05$. For an individual forcing mode (upper panel), the
tidally dissipated power is a periodic signal with period $P_{\rm orb}/(2|k-q|)$ for $k\neq q$, where
the factor of 2 in the denominator arises because dissipation scales
as the square of the velocity. 
When the planet is subject to two forcing modes with forcing frequencies $\sigma_i$ and $\sigma_j$, the flow velocities from individual modes superpose linearly, whereas the dissipation, being quadratic in the velocity, is characterized by the frequencies $2\sigma_i,2\sigma_j,\sigma_i+\sigma_j,$ and $|\sigma_i-\sigma_j|$. Since the ratios  of these frequencies are irrational, and beat frequencies $|\sigma_i\pm\sigma_j|$ are generally incommensurate with $n_{\rm orb}$, there is no single period over which the tidally dissipated power strictly repeats. Departure from periodicity increases with increasing $e_{\rm p}$ (Figure~\ref{Fig_Ts_variations}) as additional forcing modes contribute.

\item \textit{Hotspot scatter:}
In Figure~\ref{Fig_hotspot} we plot the longitude $\lambda_{\rm h}$ 
of the hottest point on the dayside (i.e.~the ``hotspot'' longitude,
sampled stroboscopically at integer multiples of the orbital period), 
together with the concurrent variation in average surface
temperature.  With only stellar insolation and no tidal heating, a
synchronously or pseudo-synchronously rotating planet would have its
hotspot located at or near the substellar point ($\lambda_{\rm h}=
0^\circ$). Lava tidal flows add a heterogeneous heating pattern that
corotates at the forcing frequencies $\sigma=q\Omega_{\rm p} - k n_{\rm
  orb}$, decoupling $\lambda_{\rm h}$ from the substellar
point. In the planet-fixed frame, the heating pattern drifts at a phase velocity of $-{\sigma }/{q}$ (see equation \ref{expansion_UT}) such that eastward drift (toward the afternoon hemisphere, or prograde motion) corresponds to $\sigma<0$, and vice versa. As $e_{\rm p}$ increases, more frequencies become relevant,
widening the range of hotspot longitudes. For  pseudo-synchronous $\Omega_{\rm p}>n_{\rm orb}$, the $(q,k)=(2,3)$ waves typically have $\sigma<0$ so that the heating pattern drifts eastward, biasing the hotspot toward the afternoon hemisphere, whereas the $(2,1)$ waves have $\sigma>0$, shifting the hotspot westward toward the morning hemisphere. The observed stroboscopic wandering of $\lambda_{\rm h}$ reflects the superposition of these and more modes. The eastward bias can be understood by the relative weights of the Hansen coefficients [e.g., $X_3^{-3,2}(e_{\rm p})/X_1^{-3,2}(e_{\rm p})=7 + \mathcal{O}(e_{\rm p}^3)$].
\end{itemize}

All the time-variable behaviors discussed above are illustrated in Figure~\ref{Fig_snapshots_Ts}, which provides heat maps of the dayside surface sampled at several periastron and apastron passages.

\section{Magma oceans in thermal equilibrium}\label{Section_thermal_equilibria} 
The previous section left the magma ocean thickness $H$ as a free parameter. Here we relate $H$ to the thermodynamics of the underlying mantle.  We will identify steady states where the heat deposited by tidal dissipation can be carried to the surface by mantle convection and ultimately radiated away. These thermal equilibria determine $H$. 

\subsection{Interior structure profiles}\label{subsec:interior_structure}
We follow \citet[][]{moore2003tidal,moore2006thermal}{}{} and search for thermal equilibrium states that balance internal heating with convective cooling, using a one-dimensional energy balance. We continue to treat our planet as airless (for studies that relax this assumption, see e.g. \citealt{zahnle2015tethered,lichtenberg2021vertically,nicholls2024magma}).

For a viscous layer heated from below and cooled from above, the onset of convection is governed by the Rayleigh-Bénard criterion \citep[e.g.][]{Chandrasekhar1961,turcotte2002geodynamics}:
\begin{equation}\label{rayleigh_nuumber}
    Ra\equiv \frac{\rho g \alpha \Delta T_{\rm sa} H^3}{\eta \kappa}> Ra_{\rm c} \approx 1708
\end{equation}
where $Ra$ is the Rayleigh number, $Ra_{\rm c}$ is the critical threshold for convection,  $\rho$ is the density, $\alpha$ the thermal expansivity, $H$ the layer's thickness, $\eta$ the dynamic viscosity, $\kappa$ the thermal diffusivity, and $\Delta T_{\rm sa}$ the superadiabatic temperature difference. The latter is the excess of the actual top-to-bottom temperature drop over the adiabatic drop, taken across the entire ocean thickness $H$. We assume that excess is localized to a thin boundary layer at the ocean surface \citep[e.g.][]{schubert2001mantle}. Thus we evaluate
\begin{equation}\label{eq:tsa}
\Delta T_{\rm sa} = T_{\rm p}-T_{\rm s}
\end{equation}
where $T_{\rm s}$ is the actual surface (blackbody) temperature (eq.~\ref{eq:Tsurfacerad}), and $T_{\rm p}$ is the surface potential temperature, defined 
as the temperature a parcel would have if brought adiabatically from the interior to the surface. Formally $T_{\rm p}$ is a free parameter; it is a proxy for the mantle adiabat which is determined not only by tidal heating but  also by radiogenic heating and the gravitational heat of formation, and the details of the planet's formation history. In section \ref{subsec:deep}, we will see that, given a planet mass, radius, and orbital period, only a certain value of $T_{\rm p}$ allows for a stable thermal equilibrium between tidal heating and convective cooling.

For $Ra \gg Ra_{\rm c}$, the bulk interior relaxes toward an adiabat, while the superadiabatic gradient is confined to boundary layers that regulate the net heat flux \citep[e.g.][]{schubert2001mantle}. Experiments constrain the dynamic viscosity of near-liquidus peridotitic melt at low pressures to be ${\sim} 0.01 {-} 0.1~{\rm Pa~s}$ \citep[e.g.][]{dingwell2004viscosity,liebske2005viscosity}{}{}, increasing to 100 Pa~s for smaller melt fractions \citep[e.g.][]{kushiro1986viscosity,rubie2003mechanisms}{}{}. For the magma ocean parameters we will determine later (section \ref{subsec:deep}), $Ra \gg Ra_{\rm c}$.


To see how a rocky planet with prescribed $(R_{\rm p},M_{\rm p})$ divides into liquid, solid, and intermediate ``mushy'' phases, we follow the treatment adopted for the Earth in \citet{farhat2025tides}, except that we rescale all functions therein from the depth domain to the pressure domain assuming hydrostatic balance. Details of the procedure are in Appendix \ref{Appendix_mantle_phase_diagram}. 
For a prescribed potential temperature $T_{\rm p}$, one obtains smooth radial profiles of temperature, pressure, and density.

\begin{figure}[t]
\includegraphics[width=.47\textwidth]{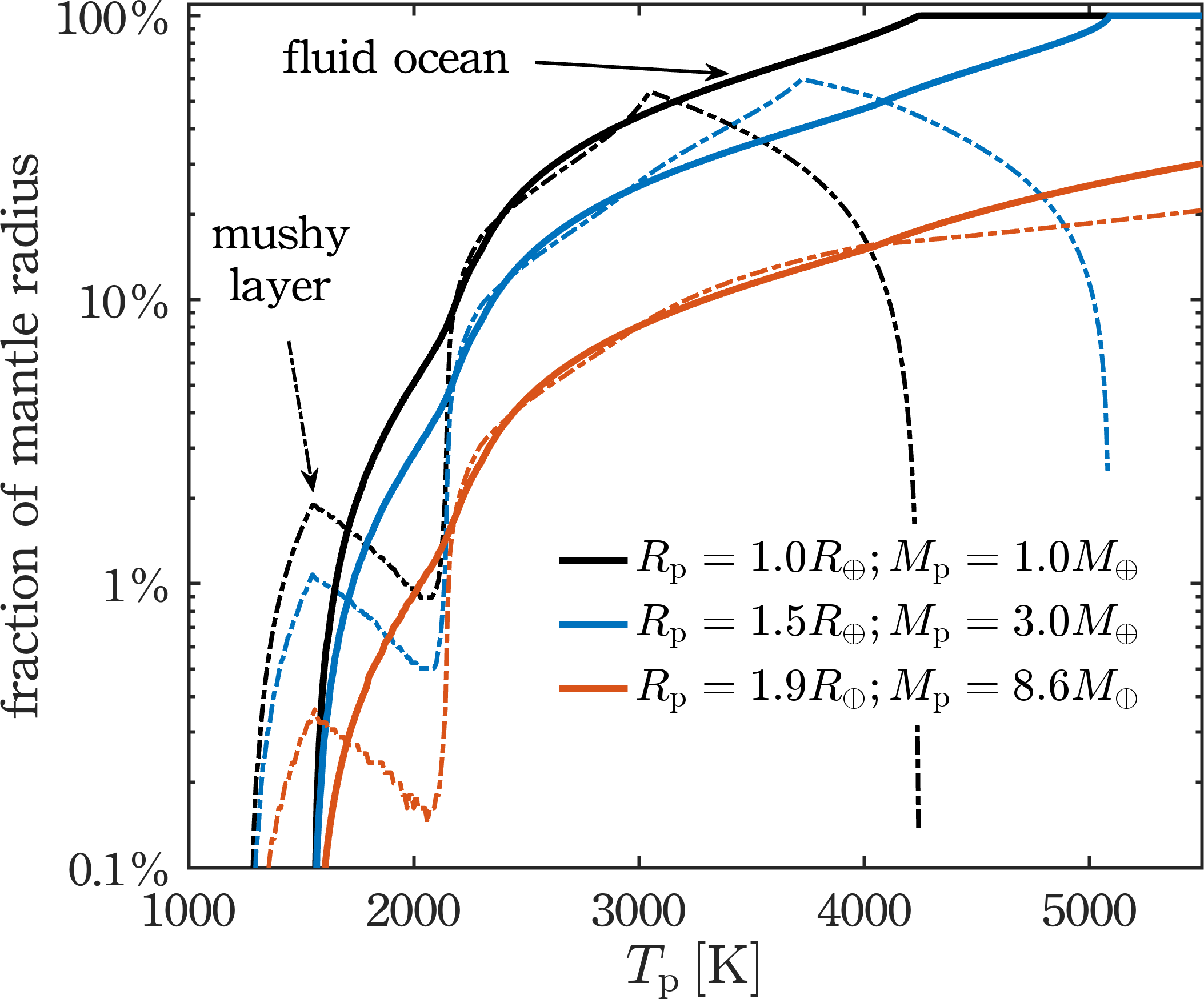}
\caption{Sample interior structures for different ($R_{\rm p}, M_{\rm p}$). Solid curves plot the thickness $H$ of the fluid magma ocean as a function of potential temperature $T_{\rm p}$,  while dashed curves plot the thickness of the mushy layer, both relative to the full thickness of the mantle. We adopt an Earth-like mantle volume fraction of $84\%$; the corresponding mantle radius is $0.46R_{\rm p}$. For calculation details, see Appendix \ref{Appendix_mantle_phase_diagram}.}
\label{Fig_ocean_thickness}
\end{figure}

Figure~\ref{Fig_ocean_thickness} shows, for three different planets, the relative thicknesses of the fluid magma ocean, intermediate mushy layer, and underlying solid mantle, as functions of $T_{\rm p}$. The base of the fluid ocean is defined by the crossing of the adiabat with the critical melt profile, $T^{\rm crit}(r)$, where $r$ is the radial depth. The mushy medium is defined by the layer between $T^{\rm crit}(r)$ and the solidus $T^{\rm sld}(r)$. Both $T^{\rm crit}$ and $T^{\rm sld}(r)$  are defined in Appendix \ref{Appendix_mantle_phase_diagram}. As shown in Figure~\ref{Fig_ocean_thickness}, with increasing $T_{\rm p}$, a mushy layer first emerges in the upper mantle and grows in size until the fluid magma ocean is established around $T_{\rm p} \sim 1600~\rm K$. At high enough $T_{\rm p}$, a planet the size and mass of the Earth would have its mantle fully molten and fluid-like (black curves). This is in contrast to a super-Earth like 55~Cancri~e, which even at $T_{\rm p}=4500~\rm K$ would still have its mantle differentiated into a viscoelastic solid at the bottom, a thick fluid magma ocean at the surface, and a mushy layer in between (orange curves).

{These interior structures are computed under the assumption that the convecting mantle is efficiently mixed and adiabatic in bulk, while the superadiabatic contrast is confined to a thin upper thermal boundary layer. We adopt this standard boundary-layer convective treatment for simplicity, although it is not a fully self-consistent model of internally heated convection. If tidal heating is the dominant heat source, a more realistic interior structure may exhibit a thermal maximum within the interior rather than at the base of the convecting layer, together with small residual thermal gradients that produce local departures from adiabaticity \citep[see e.g. Chapter 9 of ][]{jaupart2010heat}. In that case, the detailed partition between the fluid, mushy, and solid regions in Figure~\ref{Fig_ocean_thickness} could shift somewhat, but the first-order global energetics studied below are less affected by the choice of the convective model.}

\subsection{Convective flux in the tidally heated mantle}
The convective heat flux carried by the magma ocean is commonly parameterized in terms of a conductive heat flux:
\begin{equation}
    \mathcal{F}_{\rm conv} = Nu \,C_{\rm p}\, \rho\, \kappa \,\Delta T_{\rm sa} /H
\end{equation}
where $c_{\rm p}$ is the isobaric specific heat (so that $c_{\rm p} \rho \kappa$ is the thermal conductivity). The dimensionless Nusselt number $Nu > 1$ quantifies the efficiency of turbulent convection.  Two Nusselt-Rayleigh laws are widely used \citep[e.g.][]{heslot1987transitions,siggia1994high,grossmann2000scaling}:
\begin{align}
    \text{Soft turbulence (lower $Ra$):} \quad Nu &= 0.089 Ra^{1/3} \label{soft}\\
     \text{Hard turbulence (higher $Ra$):}\quad Nu &= 0.22 Ra^{2/7} Pr^{-1/7} \label{hard}
\end{align}
with $Pr = \eta/(\rho\kappa) $ denoting the Prandtl number. The former describes a regime where viscous boundary layers sandwiching the convective region are laminar, with heat transfer throttled by boundary layer conduction and plume shedding. The latter pertains to turbulent boundary layers. Equating the two prescriptions gives a crossover at
\begin{equation}
    Ra_* = \left( \frac{0.22}{0.089} Pr^{-1/7}\right)^{21} 
\end{equation}
above which hard turbulence should prevail. For silicate melts ($Pr\sim 10-10^{4}$), $Ra_*\lesssim10^6$. It is likely that $Ra > Ra_*$ (by orders of magnitude), in which case the hard turbulence law should hold. However, because the hard and soft prescriptions scale with $Ra$ very similarly, the value for $Nu$ predicted by hard turbulence is at most a factor of a few smaller than that predicted by soft turbulence. We elect to use the soft relation (\ref{soft})  as it gives larger $Nu$, i.e. stronger convective cooling, as a conservative test for whether tidal heating can sustain melt. {This choice is also conservative with respect to how the heat from tidal stresses is deposited: 
for a fixed upper thermal boundary layer contrast, internally heated convection yields a smaller cooling flux than our 
model which assumes the heat is deposited purely at the lower boundary \citep[e.g.][]{jaupart2010heat}.}

\begin{figure*}[ht]
\includegraphics[width=\textwidth]{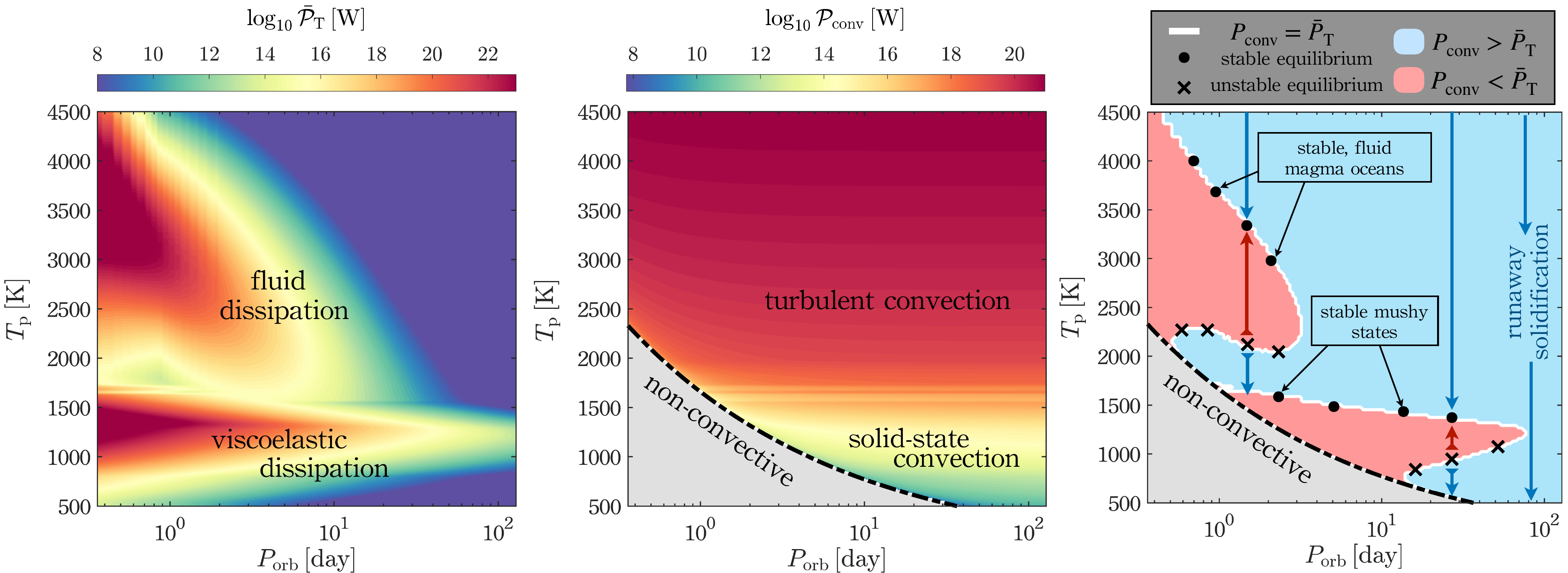}
\caption{Mantle thermodynamic equilibria in ($P_{\rm orb}-T_{\rm p})$ space for our fiducial super-Earth. We fix here $\sigma_{\rm R}=10^{-3}~{\rm s}^{-1}$ and $e_{\rm p}=0.03$. Left panel plots the tidally generated 
power, middle panel plots the convective cooling  power, 
while the third panel plots the difference between the two powers. Where the difference is zero are stable equilibria (annotated with solid circles) and unstable equilibria (crosses).}
\label{Fig_fluxes_equilibria}
\end{figure*}

If there is no surface fluid ocean, 
convective overturn proceeds by solid-state creep. The heat flux from stagnant-lid convection is given by \citep[e.g.][]{davies1988heat}:
\begin{equation}
     \mathcal{F}_{\rm conv} =0.3 \,k \,\Delta T_{\rm sa}^{4/3}\left(\frac{\alpha \rho g}{\kappa \eta_{\rm s}}\right)^{1/3}
\end{equation}
where $\eta_{\rm s}$ is the solid viscosity (see Eq.~\ref{viscosity_temperature}). 

We equate the convective power
\begin{equation}
\mathcal{P}_{\rm conv} \equiv 4\pi R_{\rm p}^2 \mathcal{F}_{\rm conv}
\end{equation}
with the tidal power $\overline{\mathcal{P}}_{\rm T}$ given by \eq{average_PT}, averaged over 10 orbital periods. The high-frequency thermal oscillations experienced by the surface (as shown in Figure~\ref{Fig_Ts_variations}) average out in the mantle. For typical forcing frequencies ($\sigma\sim10^{-4}-10^{-6}$ s$^{-1}$) and thermal diffusivities of molten silicates ($0.5-2 \times10^{-6}~{\rm m^2\,s^{-1}}$), the thermal skin depth $\sqrt{2\kappa/\sigma}$ is only a few meters; at greater depths, high frequency variations are screened out. Moreover, the convective overturn timescale is orders of magnitude longer than the orbital period. 
Convection in the magma ocean thus responds only to the secular average of tidal heating. {As we are after global energetics, we distribute this tidal heating uniformly throughout the magma ocean, since we use a single Rayleigh-drag coefficient without radial dependence. Dissipation may however be depth-dependent and could be enhanced toward the base of the ocean, where the flow interacts with the more viscous substrate underneath and heating by bottom friction is amplified. }

{In addition to dissipation in the fluid ocean, we account for dissipation in the solid and mushy portions of the mantle, both of 
which we treat with a one-phase viscoelastic model using a simplified Maxwell rheology and 
temperature-dependent viscosity (Appendix \ref{Appendix_solid_tides}). The mushy layer contributes to the thermal budget as a melt-weakened viscoelastic solid, but not as a biphasic medium with distinct rheology. The present calculation therefore does not capture additional dissipation channels that may arise in partially molten, poroviscoelastic media (see discussion in Section \ref{sec:sum}).} 

\subsection{Thermal equilibria}\label{subsec:deep}

We solve for thermal equilibria where convection  carries tidally generated heat from the mantle to the surface, whereupon it is radiated to space.  
Figure~\ref{Fig_fluxes_equilibria} summarizes results for our fiducial system. The left panel maps the tidally generated power $\overline{\mathcal{P}}_{\rm T}$ vs.~orbital period $P_{\rm orb}$ and mantle potential temperature $T_{\rm p}$. The orbital period controls the frequency and amplitude of tidal forcing, while the potential temperature defines the interior structure, specifically the thickness of the fluid magma ocean (which could be zero). 

For the interior structure profiles shown in Figure \ref{Fig_ocean_thickness}, mantles with $T_{\rm p}\lesssim1600~{\rm K}$ do not have fluid layers. Consequently, at such low $T_{\rm p}$, only solid-state tides operate (modeled according to Appendix \ref{Appendix_solid_tides}), and these give rise to a peak in dissipation around 
$T_{\rm p} \approx 1300~{\rm K}$, as seen in the bottom half of the left panel of Figure~\ref{Fig_fluxes_equilibria}. The peak is a generic feature of viscoelastic relaxation in solids \citep[e.g.][]{nowick2012anelastic,efroimsky2012tidal}.  For the Maxwell mantle considered here, the relevant timescale is 
the Maxwell time $\tau_{\rm M}=\eta (T_{\rm p})/\mu_{\rm s}$, where $\mu_{\rm s}$ is the rigidity of the mantle. Dissipation is small in the elastic limit ($\sigma\tau_{\rm M}\gg 1$) and in the viscous limit ($\sigma\tau_{\rm M}\ll 1$), and peaks in the viscoelastic regime near $\sigma\tau_{\rm M}=1$. In the elastic limit, as $T_{\rm p}$ increases, the viscosity $\eta$ and by extension the heating rate decrease exponentially (Eq.~\ref{viscosity_temperature}), in practically the same way across different rheologies (e.g.~Maxwell vs.~Andrade; \citealt{Castillo-Rogez}). In the viscous limit, dissipation also declines with increasing temperature, with quantitative differences between models (Maxwell declines faster than Andrade).

For $T_{\rm p}\gtrsim 1600~{\rm K}$, the adiabat is able to cross the critical melt 
curve and then the liquidus. A surface fluid layer is created atop a mushy layer which in turn overlies the solid mantle. The presence of fluid tides within the magma ocean causes a second peak to emerge in the left panel of Figure \ref{Fig_fluxes_equilibria}, centered on $T_{\rm p} \approx 3500$ K. This peak is the same peak seen in Figure~\ref{Fig_PT_over_Pins}. At smaller $T_{\rm p}$, the ocean is shallow and tidal dissipation is minimal; at larger $T_{\rm p}$, the forcing is too slow to move the full depth of the ocean coherently and perform dissipative work. 
The existence of the fluid-tides peak is essential to establishing fluid thermal equilibria; in particular, the high $T_{\rm p}$ (large ocean thickness) side of the peak will be seen shortly to yield stable equilibria. Varying $\sigma_{\rm R}$ from our fiducial value of $10^{-3}~{\rm s}^{-1}$ to $10^{-2}~{\rm s}^{-1}$ alters the quantitative but not qualitative features of the fluid-tides peak; note how in Fig.~\ref{Fig_PT_over_Pins} changing $\sigma_{\rm R}$ does not much change the peak value of $\overline{\mathcal{P}}_{\rm T}$. 

In the middle panel of Figure \ref{Fig_fluxes_equilibria}, we show the planet's convective luminosity, $\mathcal{P}_{\rm conv} = 4\pi R_{\rm p}^2 \mathcal{F}_{\rm conv}$. For $T_{\rm p}\lesssim1600~{\rm K}$, 
only solid-state convection is present. For $T_{\rm p}\gtrsim1600~{\rm K}$, turbulent convection is triggered in a fluid magma ocean. The latter carries orders-of-magnitude more power 
than solid-state convection. The convective power depends only weakly on $P_{\rm orb}$ via $\Delta T_{\rm sa}(T_{\rm s})$. In contrast, the dependence of $\mathcal{P}_{\rm conv}$ on $T_{\rm p}$ is much stronger since $T_{\rm p}$ enters not only through $\Delta T_{\rm sa}$ but also 
exponentially through the viscosity.  
The gray shaded region in the 
middle panel of Figure \ref{Fig_fluxes_equilibria} 
marks where the surface temperature $T_{\rm s}$ exceeds $T_{\rm p}$ and convection shuts off.

The right panel of Figure \ref{Fig_fluxes_equilibria} shows the thermal balance. Blue regions indicate net cooling, red indicates net heating, and white contours mark equilibria where $\overline{\mathcal{P}}_{\rm T}=\mathcal{P}_{\rm conv}.$ The meaning of this equality is that convection can fully carry the power deposited by tides out of the planet, thereby maintaining the planet in thermal steady state. At fixed $P_{\rm orb},$ an equilibrium is stable if an increase in $T_{\rm p}$ (away from equilibrium) results in net cooling, and if a decrease in $T_{\rm p}$ results in net heating. Otherwise the equilibrium is unstable.

For orbits with $P_{\rm orb}\gtrsim 80~{\rm days}$, no interior configuration allows tidal heating to balance convective cooling, and the mantle is destined to solidify on a viscosity-controlled timescale. Closer in, tidal heating increases and can counter convection, turning the mantle slightly mushy and producing both stable and unstable equilibria --- these outline the red triangular region in the bottom half of the right panel of Fig.~\ref{Fig_fluxes_equilibria}. 
The stable branch spans $T_{\rm p} \simeq 1250$–$1650~\rm{K}$ depending on $P_{\rm orb}$. This branch corresponds to the elastic side of the viscoelastic peak in dissipation; as $T_{\rm p}$ increases away from peak dissipation, viscosity decreases by orders of magnitude, reducing tidal heating and increasing convective cooling until heating and cooling balance.

For our fiducial super-Earth, a second set of equilibria appears at $P_{\rm orb}\lesssim 3.5~{\rm days}$. The stable branch starts at $T_{\rm p}\simeq2500~{\rm K}$ for $P_{\rm orb}\simeq3.25~{\rm days}$,  and reaches $T_{\rm p} \simeq 4050~{\rm K}$ at $P_{\rm orb}=0.7~{\rm days}$ (Fig.~\ref{Fig_fluxes_equilibria}). These equilibria indicate that once a surface magma ocean is formed, fluid-driven dissipation is sufficient to sustain the fluid melt as long as the tidal forcing persists, i.e., as long as the assumed eccentric orbit can be maintained (Figures~\ref{Fig_ep_sf} and \ref{Fig_tau_ec}, Appendix \ref{Section_maintaining_eccentricity}). 

\begin{figure}[t]
\includegraphics[width=.47\textwidth]{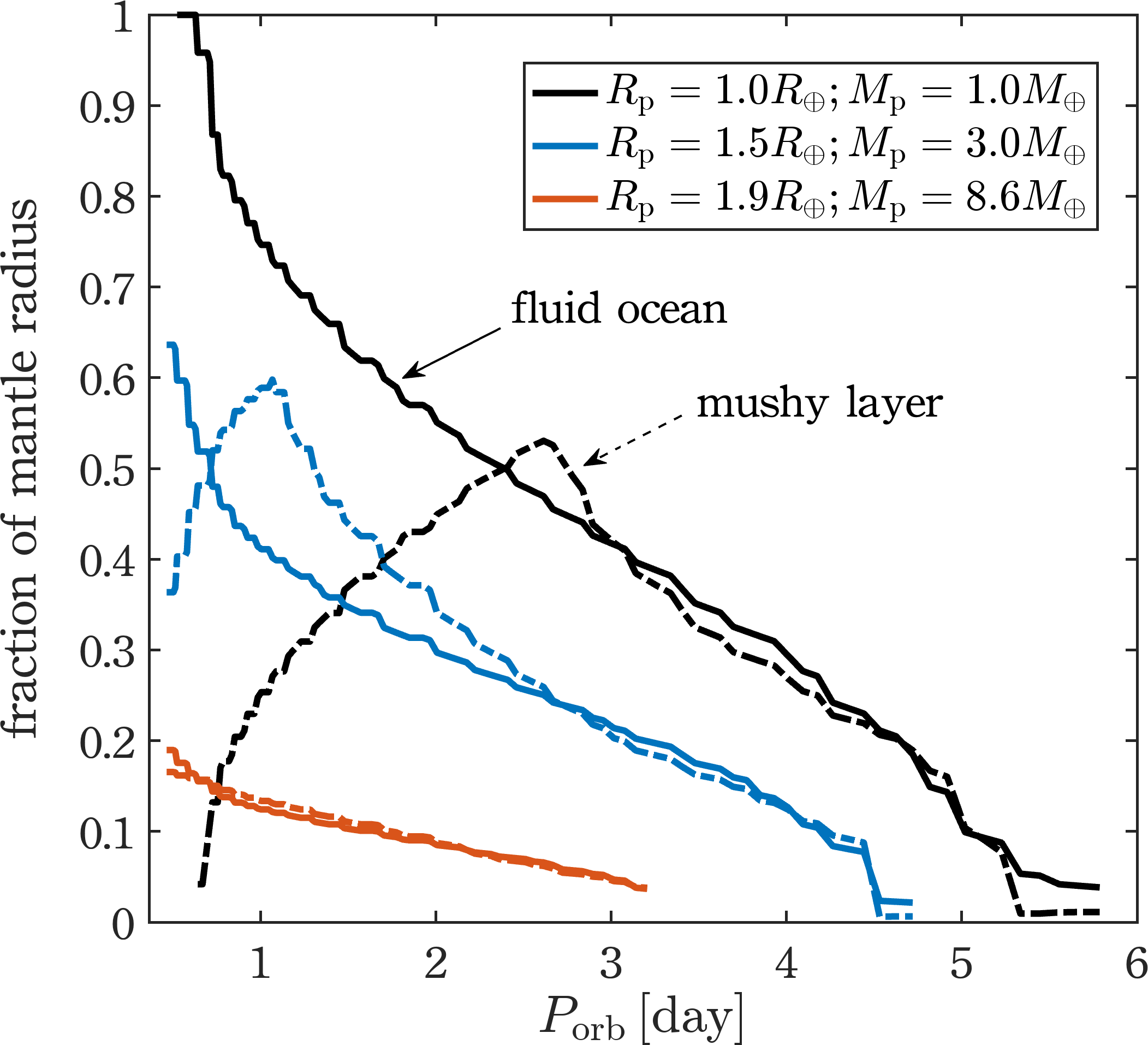}
\caption{Equilibrium interior structures of rocky planets including heating from dissipation of lava tides. Solid curves plot the equilibrium thickness of the surface magma ocean, and dashed curves the equilibrium thickness of the mushy layer underneath, both as a fraction of the mantle radius $\simeq 0.46R_{\rm p}$. In the fluid equilibrium states referred to here, the tidal power $\overline{\mathcal{P}}_{\rm T}$ is carried in full by the convective power $\mathcal{P}_{\rm conv}$ (see the high-$T_{\rm p}$ stable branch in the third panel of Fig.~\ref{Fig_fluxes_equilibria}).}
\label{Fig_H_Porb}
\end{figure}

The high-temperature equilibrium ocean thickness $H$ is plotted against orbital period $P_{\rm orb}$ in Figure~\ref{Fig_H_Porb}, together with the thickness of an underlying mushy layer, both as fractions of the mantle thickness. For our fiducial super-Earth (orange curves), the magma ocean extends from $\sim$5\% of the mantle at $P_{\rm orb} = 3.25$~d, to $\sim$20\% at $P_{\rm orb} = 0.3$~d. Similar numbers characterize the mushy layer. For this super-Earth, the bottommost portion of the mantle remains solid. For smaller, Earth-like planets (black curves), a layer of fluid melt can be sustained by tidal heating as far out as $P_{\rm orb} = 5.7$ d (and this limit would increase for higher eccentricities). At sub-day periods, an Earth-like planet (with the assumed eccentricity of $e_{\rm p}=0.03$) has a fluid ocean that extends to the core-mantle boundary.

The creation of equilibrium magma oceans does not depend on $\overline{\mathcal{P}}_{\rm T} > \overline{\mathcal{P}}_{\rm ins}$. {In fact, the equilibria identified in Figure~\ref{Fig_fluxes_equilibria} correspond to the opposite (although they need not; more on this below).} Stellar insolation provides a surface boundary condition for convection (recall \eq{eq:tsa}, $\Delta T_{\rm sa} = T_{\rm p}-T_{\rm s}$), while tides deposit heat into the interior that needs to be removed by convection to maintain steady state. Large $\overline{\mathcal{P}}_{\rm ins}$ throttles $\mathcal{P}_{\rm conv}$ down, enough for the latter to carry $\overline{\mathcal{P}}_{\rm T}$ and maintain an ocean in equilibrium. {Interior equilibria corresponding to thick oceans and satisfying $\overline{\mathcal{P}}_{\rm T} > \overline{\mathcal{P}}_{\rm ins}$ can be  obtained for larger values of $\sigma_{\rm R}$ (see Figure~\ref{Fig_PT_over_Pins}).}

\subsection{Timescales}\label{subsec:timescales}
To reach one of the stable equilibria for a given $P_{\rm orb}$, the interior evolves according to
\begin{equation}
    C_{\rm m}(T_{\rm p}) \frac{dT_{\rm p}}{dt} = \overline{\mathcal P}_{\rm T}(T_{\rm p},P_{\rm orb}) - \mathcal{P}_{\rm conv}(T_{\rm p},P_{\rm orb}),
\end{equation}
with $C_{\rm m}$ being the effective heat capacity of the actively convecting mantle, which we define to be 
\begin{equation}
 C_{\rm m}(T_{\rm p})
= 4\pi \!\int_{R_{\rm m}}^{R_{\rm p}} \rho(r) r^2 \left( \,c_{\rm p}(r) 
+ H_{\rm f}\frac{\partial F_{\rm m}}{\partial T} \right) \,dr
\end{equation}
where $R_{\rm m}=0.46 R_{\rm p}$ is the thickness of the mantle, $\rho$ is the density profile, $c_{\rm p}$ is the isobaric specific heat, $H_{\rm f}$ is the latent heat of fusion, and $F_{\rm m}$ is the melt fraction; all these quantities are defined in Appendix \ref{Appendix_mantle_phase_diagram}.  The time it takes to reach a stable equilibrium temperature $T_{\rm p;eq}$ starting from $T_{\rm p;0}$ is
\begin{equation}
\tau_{\rm settle}
=\int_{T_{\rm p;0}}^{T_{\rm p;eq}}\frac{C_{\rm m}(T_{\rm p})}{\big|\overline{\mathcal P}_{\rm T}(T_{\rm p},P_{\rm orb}) - \mathcal{P}_{\rm conv}(T_{\rm p},P_{\rm orb})\big|}\,dT_{\rm p}.
\end{equation}
Computing $\tau_{\rm settle}$ on the $(P_{\rm orb,}T_{\rm p})$ grid of Figure~\ref{Fig_fluxes_equilibria}, we find that planets with $P_{\rm orb}\lesssim3.5~{\rm days}$ settle onto the stable, high temperature equilibrium branch over $\tau_{\rm settle}\sim10^2-10^{3}~{\rm yr}$, while those with $3.5\lesssim P_{\rm orb}\lesssim80~{\rm days}$ settle onto the solid-state thermal equilibrium over $\tau_{\rm settle}\sim10^5-10^{6}~{\rm yr}$.

\begin{figure}[t]
\includegraphics[width=.47\textwidth]{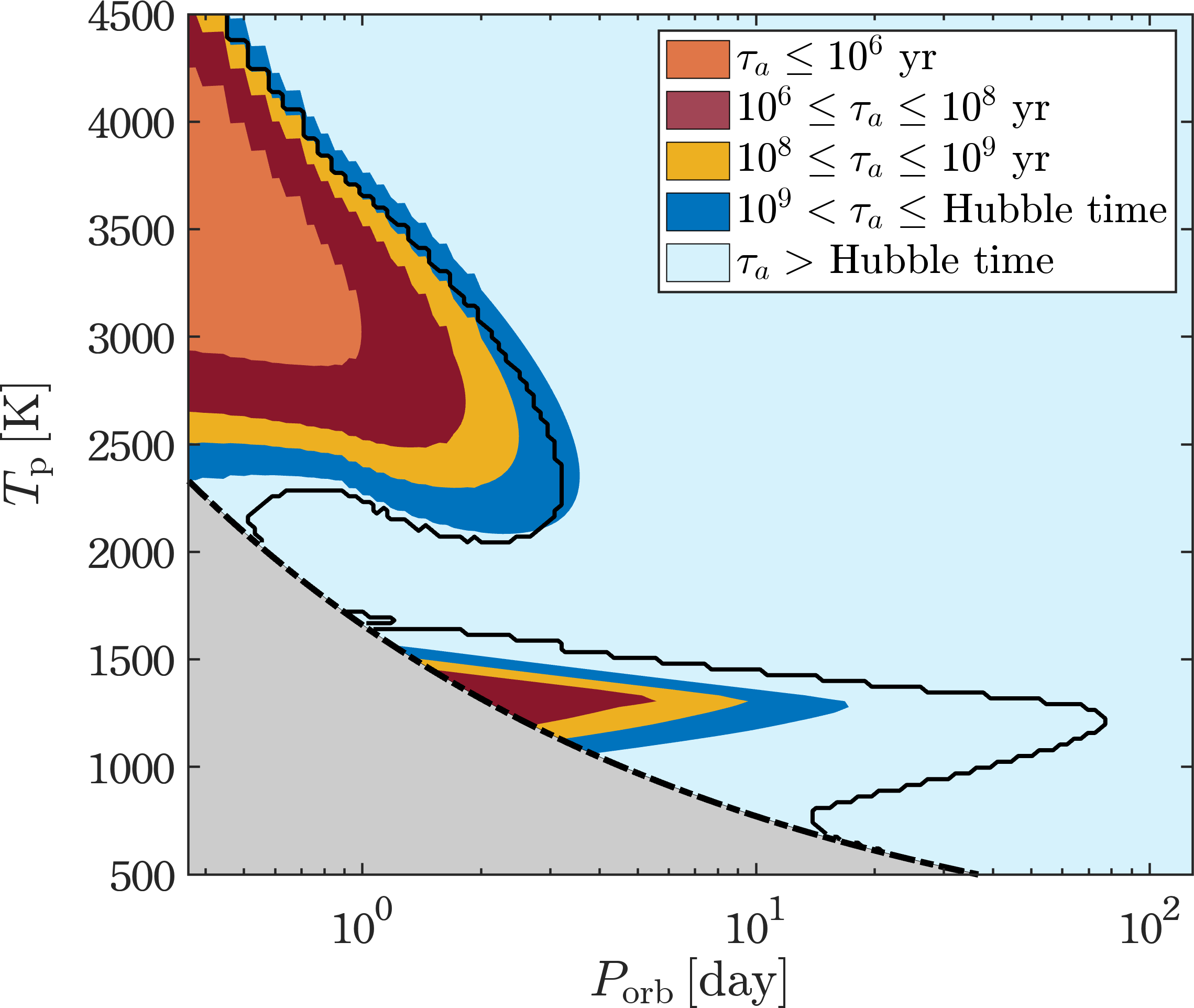}
\caption{Inspiral timescale $\tau_{\rm a}$, computed with \eq{eq_inspiral_tau} using the tidally dissipated power plotted in the first panel of Fig.~\ref{Fig_fluxes_equilibria}. Black curves trace the tidally controlled thermal equilibria identified in the third panel of Fig.~\ref{Fig_fluxes_equilibria}. Inspiral times are on the order of Gyrs, comparable to the ages of planetary systems.}
\label{Fig_inspiral_time}
\end{figure}

Once the interior reaches one of these steady states, the power needed to maintain its state originates mostly from the orbit, whose 
energy is converted into heat (a small fraction of the energy comes from the spin since the planet is not perfectly synchronous). The loss of orbital energy leads to orbital decay on the timescale
\begin{equation}\label{eq_inspiral_tau}
\tau_{\rm a}\equiv \frac{a_{\rm p}}{|\dot{a}_{\rm p}|}=\frac{GM_\star M_{\rm p}}{2a_{\rm p}\overline{\mathcal{P}}_{\rm T}} \,.
\end{equation}
Figure~\ref{Fig_inspiral_time} shows $\tau_{\rm a}$ for the values of $\overline{\mathcal{P}}_{\rm T}$ shown in Figure~\ref{Fig_fluxes_equilibria}. The black curves trace out the equilibria identified in the right panel of Figure~\ref{Fig_fluxes_equilibria}. Low-temperature, solid-state equilibria drive negligible inward migration ($\tau_{\rm a}\geq13.8 \times10^{9}~{\rm yr}$). High-$T_{\rm p}$ fluid equilibria have
$\tau_{\rm a}\simeq 10^{9}~{\rm yr}$ for $P_{\rm orb}\leq 0.7~{\rm day}$, and $10^9\leq\tau_{\rm a}\leq 10^{10}~{\rm yr}$ for $0.7\leq P_{\rm orb}\leq 3.5~{\rm days}$. 
Thus some orbital migration may be expected over the multi-Gyr ages of magma ocean worlds, especially those with the shortest periods.

\section{Summary and Outlook}\label{sec:sum}
Stellar irradiation may create ``puddles'' of melt on the dayside surfaces of short-period rocky planets. By contrast, a non-zero orbital eccentricity (or obliquity, or asynchronous rotation) provides a heat source that penetrates more deeply: gravitational tides, raised by the host star,  which repeatedly strain the planet's interior. We have investigated the possibility that tidal heating 
melts large portions of the mantle into magma oceans. The tidal response of this ocean may not be captured by the usual
theories of viscoelasticity. Instead, the fluid ocean may support long-wavelength (shallow-water) gravity waves that can propagate across the dayside while being damped by viscosity and friction. The dissipation of such ``lava tidal waves'' can lead to potentially observable, time-variable thermal emission.  

Eccentricity tides depend on eccentricity. For practically all known short-period rocky 
exoplanets, eccentricities measured from stellar radial velocities are
$\lesssim 0.05$--0.1. But 
eccentricities of a few percent suffice for tidal heating to
qualitatively alter the global thermodynamics. Plausibly, in a subset of systems, eccentricities of this order may be secularly forced and
maintained over timescales comparable to 
the system age, by exterior companions with measurably
eccentric orbits (Figures~\ref{Fig_ep_sf} and \ref{Fig_tau_ec}, Appendix \ref{Section_maintaining_eccentricity}).

Our solutions to the Laplace tidal equations for
creep flow across the dayside hemisphere, coupled to interior models for fluid and solid-state convection, taught us the following:
\begin{itemize}
     \item[\textit{i})] The larger the orbital eccentricity, the
       broader the spectrum of lava waves that are driven.  At
       eccentricities of a few percent, wave interference causes the
       dayside-averaged brightness temperature to vary aperiodically, on timescales faster than the orbital period by factors of 2--4 (Figures~\ref{Fig_Ts_variations} and \ref{indiv_modes}). The dayside hotspot wanders, mostly eastward of the substellar point but sometimes westward  (Figure~\ref{Fig_hotspot}). The amplitudes of these variations may be observable --- of order hundreds of K for the dayside-averaged brightness temperature, and tens of degrees for the hotspot longitude --- if the tidal power exceeds the stellar insolation in thermal equilibrium, a condition that depends on model details.

     \item[\textit{ii})] At orbital periods shorter than $\sim$3 days and eccentricities of a few percent, tidal heating creates deep magma oceans. For smaller Earth-sized planets, the fluid lava ocean can extend all the way to the
     core-mantle boundary. For larger super-Earths, the ocean, still hundreds of kilometers deep, overlies a mushy two-phase layer that transitions to a 
     solid viscoelastic lower mantle (Figure
     \ref{Fig_H_Porb}). These interior profiles represent
     stable, self-consistent thermal equilibria where the energy
     deposited into the planet interior by tides (plus radioactivity and the
     heat of formation) is transported steadily out of the mantle by a combination of fluid and solid-state convection (capped by stellar insolation as a boundary condition; Figure \ref{Fig_fluxes_equilibria}). Fluid and semi-fluid equilibria can persist against orbital (semimajor axis) decay for times comparable
     to or longer than the system age (Figure \ref{Fig_inspiral_time}), assuming eccentricity can
     be maintained by external forcing (Figure \ref{Fig_tau_ec}).
\end{itemize}

Failing eccentricity, other sources of tidal heating are planetary obliquity and asynchronous rotation (e.g.~\citealt{makarov2012conditions,correia2019spin,millholland2019obliquity}). 
{The qualitative behaviors we uncovered should not depend on eccentricity specifically, but rather on the presence of a persistent time-varying tidal potential. Irrespective of its source, this time-variable forcing can excite motions in the magma ocean and generate heat. If sufficiently strong, such heating can help sustain deep fluid oceans, and if several modes are excited simultaneously, can also produce temporally irregular surface thermal emission.}

Are lava waves driven by eccentricity tides a root cause for the dramatic time variability observed in 55 Cnc e (see Introduction)? The current upper limit of $e_{\rm p} \lesssim 0.05$ \citep[][]{bourrier201855} does not rule out significant tidal heating, but there is no clear mechanism (read: exterior perturber) by which a percent-level eccentricity for 55 Cnc e can be maintained (\citealt{ferraz2025tidal}). More accurate eccentricity measurements for more short-period planets, together with synoptic studies, are welcome.

Other proposed explanations for time variability in ultra-short period planets like 55 Cnc e invoke climate feedbacks from silicate cloud cycles (\citealt{loftus2025extreme}; see also \citealt{bromley2023chaotic}), stochastic degassing from the planet interior \citep{heng2023transient}, and 
circumplanetary/circumstellar dust/gas tori fed by planetary outflows 
\citep[][]{valdes2023investigating,patel2024jwst}. These proposals, which involve surface/atmosphere dynamics in one way or another, are not incompatible with forcing by lava tides. As equilibrium vapor pressures are exponentially sensitive to surface temperature, an irregularly heated magma ocean may drive large  aperiodic variations in atmospheric mass. The chemistry and elemental exchange between ocean and atmosphere are expected to change continuously as regions of active degassing and condensation shift across the planet.  Tidally driven thermal pulses may also power vertical flows and atmospheric escape.
{A growing number of planets are now observed to support secondary atmospheres outgassed from the underlying magma ocean \citep[e.g.][]{hu2024secondary,teske2025thick,nicholls2025volatile}.}

{An important limitation of our tidal model concerns how we treated the mushy medium between the surface fluid ocean and the lower solid mantle as a melt-weakened viscoelastic solid, rather than a  biphasic one. A more complete treatment would require a poroviscoelastic formalism that couples the deformation of the solid skeleton of the mush, pore fluid flow, and self-gravity. Recent works have made progress in this direction \citep[e.g.][]{liao2020heat,kamata2023poroviscoelastic,hay2025poro}, resolving additional dissipation channels arising from melt segregation, Darcy flow, and solid matrix compaction. In particular, while focused on Io, \citet[][]{hay2025poro} found that the quantitative importance of these additional channels is highly parameter dependent and does not generically dominate the contribution of solid state shear. Working out these issues for lava planets is an important goal for future work.}

Our assumption that the planet nightside is solid (effectively frozen) may be questioned. Day and night hemispheres are generally equally stressed by tides, and we have shown that depending on the magma ocean thickness, tidal heating may overpower stellar insolation. Heat may also be transported across the terminator by 
atmospheric flows \citep[e.g.][]{castan2011atmospheres}; the pressure head of the dayside ocean percolating melt through a permeable nightside \citep[e.g.][]{bear1972dynamics,turcotte2002geodynamics}; or horizontal convection in the planet interior \citep[e.g.][]{hughes2008horizontal,noto2025convective}. Day-night thermal contrasts as probed by phase curves may thus be muted by tidal heating.

\begin{acknowledgements}
We thank the referee Charles-Édouard Boukaré for a careful and thorough revision that improved the paper considerably. We thank Jihad Touma for insights into secular theory, Pierre Auclair-Desrotour for pointing us to the work of \citet{unno1979nonradial} and the effects of boundary conditions on tidal flows, Tim Lichtenberg for a discussion of 55~Cancri~e, and Adrien Leleu for an assessment of eccentricity measurements of close-in exoplanets. Participants of the 2025 Exoplanets and Planet Formation Conference hosted by the T.D.~Lee Institute and Shanghai Jiao Tong University provided helpful feedback.  M.F.~and E.C.~are supported by the Miller Institute for Basic Research in Science, University of California Berkeley. E.C.~acknowledges a Simons Investigator grant. 
\end{acknowledgements}

\appendix 
\section{Maintaining eccentricity by secular forcing}
\label{Section_maintaining_eccentricity}
We discuss how the inner planet's nonzero eccentricity, $e_{\rm p}\neq0$, may be maintained by secular forcing from an exterior eccentric planet. We define the normalized angular momentum vector $\vec j_{\rm p} = \sqrt{1-e_{\rm p}^2}~\hat{n}_{\rm p}$ and the Lenz vector $\vec{e}_{\rm p}=e_{\rm p}\hat{u}_{\rm p}$, where $\hat{n}_{\rm p}$ and $\hat{u}_{\rm p}$ are unit vectors pointing in the directions of the 
orbital normal and periapse, respectively. Analogous vectors with subscript `c' refer to the orbit of the exterior companion of mass $M_{\rm c}$, semi-major axis $a_{\rm c}$, and eccentricity $e_{\rm c}$. 
The orbit-averaged multipolar interaction between two planets in terms of these vectors is well documented in  studies of hierarchical triples
\citep[e.g.][]{Tremaine2009,correia2011tidal,hamers2020secular,farhat2021laplace,farhat2023case}. The secular Hamiltonian is 
\begin{equation}\label{hamiltonian1}
    H_{\rm S} = \frac{GM_{\rm c}\mu_{\rm p}}{a_{\rm c}}\left(\Psi_{\rm quad} + \Psi_{\rm oct}\right) + \mathcal{O}(\alpha^4)
\end{equation}
where we have dropped the constant Keplerian terms, $\mu_{\rm
  p}=M_{\rm p}M_{\star}/(M_\star + M_{\rm p})$ is the reduced mass of
the inner planet with $M_{\rm p}$ the inner planet's mass and $M_{\rm
  \star}$ the star's mass, $\alpha= a_{\rm p}/a_{\rm c}$, and the quadrupolar and octupolar terms of the disturbing function (assuming coplanar orbits) read as
\begin{equation}
    \Psi_{\rm quad}= -\frac{3}{4}\varepsilon_{\rm q} e_{\rm p}^2
\end{equation}
\begin{equation}
    \Psi_{\rm oct}= -\frac{15}{16}\varepsilon_{\rm q}\varepsilon_{\rm o} e_{\rm p}\cos\Delta\varpi + \mathcal{O}(e_{\rm p}^2)
\end{equation}
with $\Delta\varpi=\varpi_{\rm c}-\varpi_{\rm p}$ equal to the difference in periapse longitudes between the outer and inner orbits, and 
the dimensionless coefficients 
\begin{equation}
    \varepsilon_{\rm q} =\alpha^2 (1-e_{\rm c}^2)^{-3/2},\hspace{.5cm} \varepsilon_{\rm o} = \alpha\frac{M_{\rm \star }-M_{\rm p}}{M_{\rm \star }+M_{\rm p}} \frac{e_{\rm c}}{1-e_{\rm c}^2}.
\end{equation}
The equations governing the eccentricity evolution of each orbit are
\begin{equation}\label{milankovic_equations}
\frac{d\vec e_k}{dt}=-L_k^{-1}\left(\vec j_k\times\nabla_{\vec e_k} H_{\rm S}+\vec e_k\times\nabla_{\vec j_k} H_{\rm S}\right)-\gamma_{\rm p}\vec e_{\rm p}\delta_{k{\rm p}},
\end{equation}
with $k$ = `p' or `c', $\delta_{k{\rm p}}$ the Kronecker delta, $L_{\rm p}=\mu_{\rm p}\sqrt{G(M_\star+M_{\rm p})a_{\rm p}(1-e_{\rm p}^2)},$ and $ 
L_{\rm c}=\mu_{\rm c}\sqrt{G(M_\star+M_{\rm c})a_{\rm c}(1-e_{\rm c}^2)}$ with $\mu_{\rm c}=M_\star M_{\rm c}/(M_\star+M_{\rm c})$ the reduced mass of the companion. The last term models the tidal damping of the inner planet (ignored for the outer planet) characterized by the rate $\gamma_{\rm p}$. The latter can be estimated in our model by assuming, to first order, that tidal evolution of the inner planet proceeds at fixed orbital angular momentum, i.e., $a_{\rm p}(1-e_{\rm p}^2)=\rm constant,$ which gives 
\begin{equation}\label{eq_gamma_p}
    \gamma_{\rm p}=\frac{1-e_{\rm p}^2}{2e_{\rm p}^2 \tau_{\rm a}},
\end{equation}
with $\tau_{\rm a}$ given by \eq{eq_inspiral_tau}.

Considering $\Psi_{\rm quad}$ only, and taking the appropriate derivatives of $H_{\rm S}$, we find:
\begin{align}\nonumber
   \frac{d\vec{e}_{\rm p}}{dt}\Bigg|_{\rm quad} &= \frac{3}{2}n_{\rm p} \frac{M_{\rm c}}{M_\star+M_{\rm p}}\alpha^3 (1-e_{\rm c}^2)^{-3/2}
 \vec j_{\rm p}\cross \vec e_{\rm p} \\
 &\equiv \dot{\varpi}_{\rm p;quad}~\vec j_{\rm p}\cross \vec e_{\rm p} , \label{eq:quadprec}
 \end{align}
where $n_{\rm p}$ is the mean motion of the inner planet. Equation (\ref{eq:quadprec}) describes pure in-plane rotation of $\vec e_{\rm p}$: the companion's axisymmetric potential 
drives prograde apsidal precession of the inner orbit with angular velocity $\dot{\varpi}_{\rm p;quad}$, leaving the magnitude $e_{\rm p}$ unchanged. The companion, with mean motion $n_{\rm c}$, precesses under the inner quadrupolar torque following
\begin{equation}
\frac{d\vec e_{\rm c}}{dt}\Big|_{\rm quad}=\dot\varpi_{\rm c;quad}\,\vec j_{\rm c}\times\vec e_{\rm c},
\end{equation}
with a  rate given by
\begin{equation}
\label{varpi_c_quad}
\dot\varpi_{\rm c;quad} = \frac{3}{4}\,n_{\rm c}\,\frac{M_{\rm p}}{M_\star+M_{\rm c}}\,
\alpha^{2} + \mathcal{O}(e_{\rm p}^2).
\end{equation}
Considering the octupolar term to first order in $e_{\rm p}$, we have
\begin{equation}
   \frac{d\vec{e}_{\rm p}}{dt}\Bigg|_{\rm oct} = -\frac{15}{16}\frac{GM_{\rm c}}{a_{\rm c}}\varepsilon_{\rm q}\varepsilon_{\rm o}L_{\rm p}^{-1}
 \vec j_{\rm p}\cross \vec e_{\rm c}\equiv \vec S,
 \end{equation}
an in-plane vector  perpendicular to $\vec{e}_{\rm c}$ of magnitude  
\begin{equation}
    {s}= |\vec{S}| = \frac{5}{4}\varepsilon_{\rm o}\dot{\varpi}_{\rm p;quad},
\end{equation}
and for the outer companion, to leading order in $e_{\rm c}$,
\begin{equation}
\frac{d\vec e_{\rm c}}{dt}\Big|_{\rm oct} =-\frac{s}{e_{\rm c}}\,\frac{L_{\rm p}}{L_{\rm c}}\,\vec j_{\rm c}\times\vec e_{\rm p}.
\end{equation}
The octupole therefore introduces no new precession. Instead, it supports a linear exchange of angular momentum between the two orbits. 
 
Additional sources of apsidal precession for the inner orbit, not caused by the exterior planet, include tides raised on the planet by the star \citep[e.g.][]{kopal1972tidal,hut1981tidal}, with the precession frequency
\begin{equation}\label{tidal_precession}
   \dot{\varpi}_{\rm p;tide}  = \frac{15}{2}\,n_{\rm orb}\,\frac{M_\star}{M_{\rm p}}
   \left(\frac{R_{\rm p}}{a_{\rm p}}\right)^{5}
   \frac{1+\tfrac{3}{2} e_{\rm p}^{2}+\tfrac{1}{8}e_{\rm p}^{4}}{(1-e_{\rm p}^{2})^{5}}\,
   \mathfrak{Re}[k_{2,2}],
\end{equation}
where $\mathfrak{Re}[k_{2,2}]$ is the real part of the degree-2,
order-2 Love number (the counterpart $\mathfrak{Im}[k_{2,2}]$ drives
tidal migration and circularization). The Love number can be computed
by solving for the tidal displacement field $\zeta(\theta,\lambda,t)$
defined in \eq{eq.zetasol}, and recovering the tidal distortion
potential of the planet. It suffices here to set $\mathfrak{Re}\{k_{2,2}\}=1.5$, the cap value of the equilibrium tide for a homogeneous fluid.  

In addition to tides, the first-order, post-Newtonian, general-relativistic interaction between the star and the planet also induces apsidal precession with frequency \citep[e.g.][]{eggleton2001orbital} 
\begin{equation}\label{GR_precession}
    \dot\varpi_{\rm p;GR} = \frac{3GM_\star} {c^2a_{\rm p}}\frac{ n_{\rm p}}{1-e_{\rm p}^2},
\end{equation}
where $c$ is the speed of light. Adding effects together, we have: $\dot{\varpi}_{\rm p} =\dot{\varpi}_{\rm p;quad} + \dot{\varpi}_{\rm p;GR} + \dot{\varpi}_{\rm p;tide}$ and $\dot\varpi_{\rm c}\equiv \dot{\varpi}_{\rm c;quad}$.

Consider a fixed right-handed basis $(\hat{\boldsymbol x},\hat{\boldsymbol y},\hat{\boldsymbol z})$ with $\hat{\boldsymbol z}=\hat n_{\rm p}=\hat n_{\rm c}$, such that $\vec e_k=e_{kx}\hat{\boldsymbol x}+e_{ky}\hat{\boldsymbol y}$. The vector equations (\ref{milankovic_equations}) are resolved in two dimensions as:
\begin{align}
\dot e_{{\rm p}x}&=\,-\, \dot\varpi_{\rm p}\,e_{{\rm p}y}\;+\;\frac{s}{e_{\rm c}}\,e_{{\rm c}y}\;-\;\gamma_{\rm p} e_{{\rm p}x},\\
\dot e_{{\rm p}y}&=\;\;\, \dot\varpi_{\rm p}\,e_{{\rm p}x}\;-\;\frac{s}{e_{\rm c}}\,e_{{\rm c}x}\;-\;\gamma_{\rm p} e_{{\rm p}y},\\[2pt]
\dot e_{{\rm c}x}&=\,-\, \dot\varpi_{\rm c}\,e_{{\rm c}y}\;+\;\frac{s}{e_{\rm c}}\frac{L_{\rm p}}{L_{\rm c}}\,e_{{\rm p}y}, \\
\dot e_{{\rm c}y}&=\;\;\, \dot\varpi_{\rm c}\,e_{{\rm c}x}\;-\;\frac{s}{e_{\rm c}}\frac{L_{\rm p}}{L_{\rm c}}\,e_{{\rm p}x}.
\end{align}
Using the complex Poincaré eccentricity variables $
z_{\rm p}\equiv e_{{\rm p}x}+i e_{{\rm p}y}=e_{\rm p}e^{i\varpi_{\rm
    p}}$ and $z_{\rm c}\equiv e_{{\rm c}x}+i e_{{\rm c}y}=e_{\rm c}e^{i\varpi_{\rm c}}$, the evolution of the system is recast as:
\begin{align}
    &\dot z_{\rm p}-i\, \dot\varpi_{\rm p}\,z_{\rm p}+\gamma_{\rm p} z_{\rm p}= -\,i\,\frac{s}{e_{\rm c}}\,z_{\rm c}, \label{eq:inner_zdot}\\
    &\dot z_{\rm c}-i\, \dot\varpi_{\rm c}\,z_{\rm c}= -\,i\,\frac{s}{e_{\rm c}}\frac{L_{\rm p}}{L_{\rm c}}\,z_{\rm p}. \label{eq:outer_zdot}
\end{align}
Let each orbit have a free and a forced mode ($z_k = z_k^{(0)}+z_k^{(\rm forced)}$). We presume the free mode of the inner orbit $z_{\rm p}^{(0)}$ is damped to zero by tides. For stationary solutions ($\dot z_{\rm p}=\dot z_{\rm c}=0$), equation (\ref{eq:outer_zdot}) gives
\begin{equation}
z_{\rm c}^{(\rm forced)}
=\frac{s}{\dot\varpi_{\rm c}\,e_{\rm c}}\frac{L_{\rm p}}{L_{\rm c}}\,z_{\rm p}.
\label{eq:zc_forced_damped}
\end{equation}
Substituting $z_{\rm c}=z_{\rm c}^{(0)}+z_{\rm c}^{(\rm forced)}$ into (\ref{eq:inner_zdot}) with $\dot z_{\rm p} = 0$ gives
\begin{equation}
    \Big(\gamma_{\rm p}-i\,\dot\varpi_{\rm p}+\frac{s^{2}}{e_{\rm c}^{2}}\frac{L_{\rm p}}{L_{\rm c}}\frac{1}{-\,i\,\dot\varpi_{\rm c}}\Big)z_{\rm p}
= -\,i\,\frac{s}{e_{\rm c}}\,z_{\rm c}^{(0)}.
\end{equation}
Hence the complex forced solution is
\begin{equation}\label{eq_zp_forced}
z_{\rm p}^{(\rm forced)}=
\frac{-\,i\,\dfrac{s}{e_{\rm c}}}{\displaystyle \gamma_{\rm p}-i\,\dot\varpi_{\rm p}
+\frac{1}{-i\dot\varpi_{\rm c}}\frac{s^{2}}{e_{\rm c}^{2}}\frac{L_{\rm p}}{L_{\rm c}}\,
}z_{\rm c}^{(0)}.
\end{equation}
Taking the modulus, and taking hereafter $e_{\rm c}\approx e_{\rm c}^{(0)}$ yields the inner forced eccentricity
\begin{equation}\label{forced_ecc_general}
e_{\rm p}^{(\rm forced)}=
\frac{s}{\left|\,
\gamma_{\rm p}-i\,\dot\varpi_{\rm p}
+\dfrac{1}{-i\dot\varpi_{\rm c}}\dfrac{s^{2}}{(e_{\rm c}^{(0)})^{2}}\dfrac{L_{\rm p}}{L_{\rm c}}\,\right|}.
\end{equation}
In the limit of negligible back-reaction $(L_{\rm p}/ L_{\rm c}\simeq (M_{\rm p}/M_{\rm c})\sqrt{a_{\rm p}/a_{\rm c}}\rightarrow0)$, \eq{forced_ecc_general} reduces to
\begin{equation} \label{eq:eforcedp}
    ~e_{\rm p}^{(\rm forced)}\simeq \dfrac{s}{\sqrt{\dot\varpi_{\rm p}^{2}+\gamma_{\rm p}^{2}}}.
\end{equation}
The inner-planet eccentricity increases from the outer-planet octupolar potential through $s$, and decreases from precession and tidal damping.

Since the forced $e_{\rm p}$ enters on both sides of
(\ref{forced_ecc_general}), we solve for it by fixed point iteration,
starting with $e_{\rm p}=(5/4) \varepsilon_{\rm o}$ (the solution
without extra precession, damping, or backreaction).  Sample solutions
for $e_{\rm p}$ 
for our fiducial super-Earth are plotted as colored contours in
Figure~\ref{Fig_ep_sf}, for either $M_{\rm c} = 10 M_\oplus$ or
$M_{\rm c} = 1 M_{\rm J}$, with a fixed damping rate $\gamma_{\rm
  p}=2\times10^{-7}$ yr$^{-1}$. The latter, estimated from
\eq{eq_gamma_p} with $e_{\rm p}=0.05$ and $\tau_{\rm a}=10^9~$yr
(Figure~\ref{Fig_inspiral_time}),
is generally smaller than
$\dot\varpi_{\rm p}$.

In Figure~\ref{Fig_ep_sf}, for $\alpha \gtrsim 0.2$ and not near a mean-motion resonance, the multipole series converges slower than in more hierarchical systems. To first order in $e_{\rm p},$ the hexadecapole ($\propto \alpha^4$) affects the precession of $\vec{e}_{\rm p}$, not its magnitude, and we can estimate its correction to be on the order of $\delta\dot{\varpi}_{\rm p;hex}\sim \alpha^2 (1-e_{\rm c}^2)^{-2}\dot{\varpi}_{\rm p;quad}$, yielding a $4\%$ correction to the plotted values for $\alpha = 0.2$ and $e_{\rm c}=0.2$.

\begin{figure}[t]
\includegraphics[width=.47\textwidth]{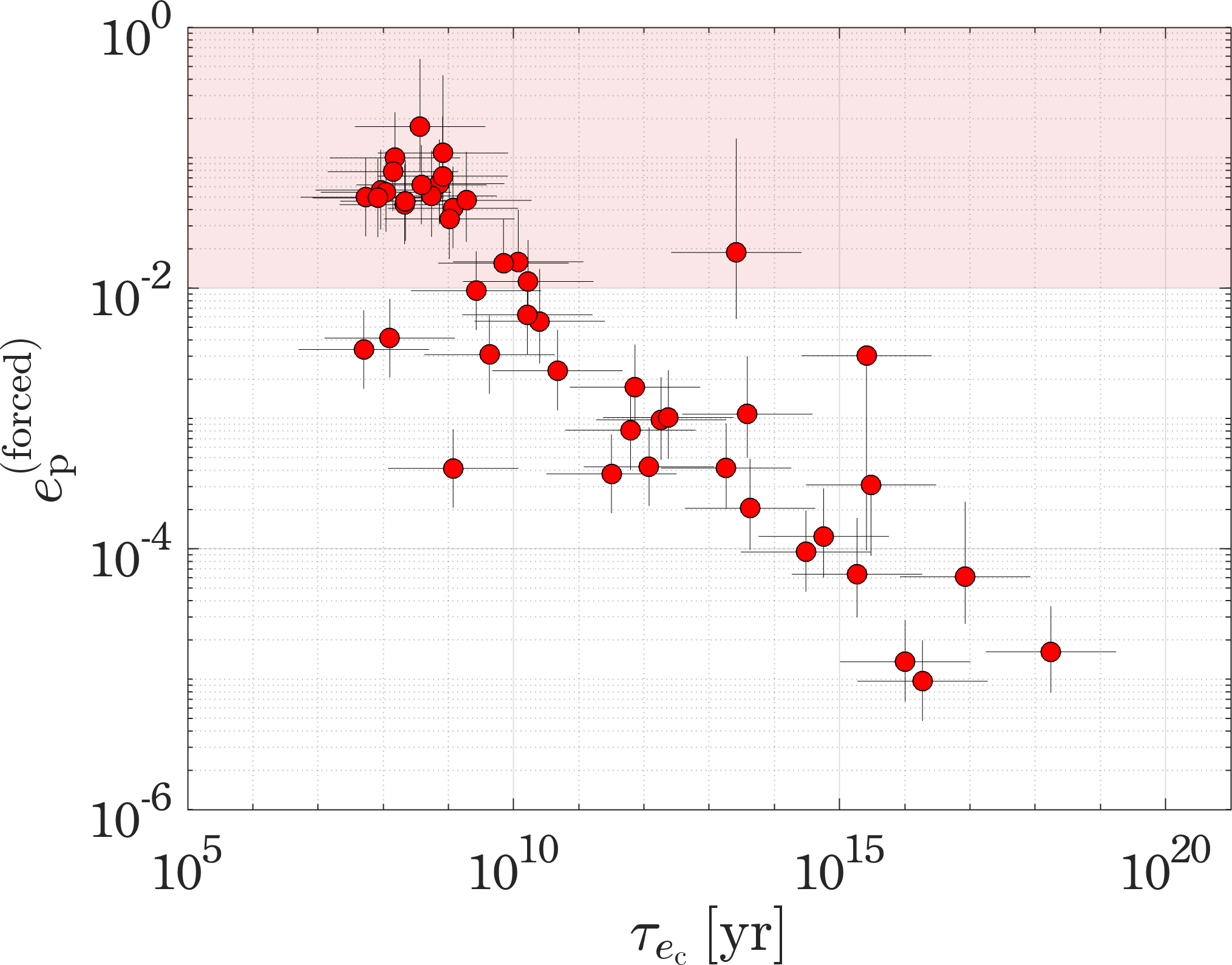}
\caption{Plotted using the sub-sample of observed short-period rocky planets (defined in section \ref{sec:intro}) known to have outer planetary companions --- the timescale $\tau_{e_{\rm c}}$ for the outer companion's eccentricity to decay, when the two planets interact secularly and the inner planet has an eccentricity $e_{\rm p}^{\rm (forced)}$ that tidally damps at rate $\gamma_{\rm p}$. Vertical error bars reflect  uncertainties in the companions' observed eccentricities $e_{\rm c}$, and horizontal error bars correspond to varying $\gamma_{\rm p}$ from $10^{-8}~{\rm s}^{-1}$ to $10^{-6}~{\rm s}^{-1}$. These $\gamma_{\rm p}$ values are calculated from \eq{eq_gamma_p} and inspired by our thermal model which predicts $\tau_a \sim 10^9$--$10^{10}$ yr for $e_{\rm p} \sim 0.03$ (Fig.~\ref{Fig_inspiral_time}). According to this figure, some rocky planets can have secularly forced eccentricities $e_{\rm p}^{({\rm forced})}$ of a few percent that last on the order of Gyrs.
} 
\label{Fig_tau_ec}
\end{figure}

The quasi-equilibrium of \eq{forced_ecc_general} is maintained as long as the outer planet is eccentric. We now estimate the timescale for circularizing the outer orbit. We first define the Angular Momentum Deficit (AMD) of orbit $k$ as \citep[][]{Laskar2000,laskar2017amd}
\begin{equation}
    {\rm AMD}_k \equiv \Lambda_k -L_k = \Lambda_k\left(1-\sqrt{1-e_k^2}\right) 
\end{equation}
where the circular-orbit angular momentum 
\begin{equation}
    \Lambda_k= \mu_k \sqrt{G(M_\star + M_k)a_k}.
\end{equation}
The only way the total AMD of the system can change is through dissipation, i.e.~tides. 
When tides act only on the inner planet, the time derivative of the total AMD of the two-planet system is
\begin{equation}\label{eq_AMD1}
    \frac{d(\rm AMD)}{dt} = \left(1-\sqrt{1-e_{\rm p}^2}\right) \dot{\Lambda}_{\rm p} - \Lambda_{\rm p} 
    \frac{e_{\rm p}^2}{\sqrt{1-e_{\rm p}^2}}\gamma_{\rm p}.
\end{equation}
Setting $\dot{\Lambda}_{\rm c} = 0$, and recognizing that $e_{\rm p}$ is dependent on $e_{\rm c}$ in the quasi-equilibrium of (\ref{forced_ecc_general}), we write 
\begin{equation}
     \frac{d(\rm AMD)}{dt} =  \frac{d(\rm AMD)}{d e_{\rm c}}\frac{de_{\rm c}}{dt}
\end{equation}
which when equated to (\ref{eq_AMD1}) yields the non-dissipative evolution of the companion's eccentricity 
\begin{equation}\label{decdt_equation}
    \frac{d e_{\rm c}}{dt} = \frac{\left(1-\sqrt{1-e_{\rm p}^2}\right)
      \dot{\Lambda}_{\rm p} - \Lambda_{\rm p} \dfrac{e_{\rm
          p}^2}{\sqrt{1-e_{\rm p}^2}}\gamma_{\rm p}}{\Lambda_{\rm p}\dfrac{e_{\rm p}}{\sqrt{1-e_{\rm p}^2}}\left(|\mathcal{K}| + e_{\rm c} \dfrac{d|\mathcal{K}|}{d e_{\rm c}}\right)+ \Lambda_{\rm c}\dfrac{e_{\rm c}}{\sqrt{1-e_{\rm c}^2}}}.
\end{equation}
Here we have used \eq{eq_zp_forced} to define the complex response $\mathcal{K}$  as $z_{\rm p}^{(\rm forced)}=\mathcal{K}z_{\rm c}^{(\rm 0)}$, so that $e_{\rm p}^{(\rm forced)}=|\mathcal{K}|e_{\rm c}^{(\rm 0)}$. Since migration is much slower than eccentricity damping, we set $\dot\Lambda_{\rm p}=0$. 
Approximating $e_{\rm p}\sim e_{\rm c}\ll1$, \eq{decdt_equation} reduces to
\begin{equation}
    \frac{de_{\rm c}}{dt}\simeq -\dfrac{L_{\rm p}\gamma_{\rm p}|\mathcal{K}|^2}{L_{\rm c}+L_{\rm p}|\mathcal{K}|^2}e_{\rm c}
\end{equation}
which gives the outer-orbit circularization timescale
\begin{equation}\label{eq_tau_ec}
    \tau_{e_{\rm c}}\equiv \left| \frac{e_{\rm c}}{\dot{e}_{\rm c}}\right|\simeq\dfrac{L_{\rm c}/L_{\rm p}+|\mathcal{K}|^2}{|\mathcal{K}|^2} \gamma_{\rm p}^{-1}.
\end{equation}
For our sample of observed planetary pairs in Figure~\ref{Fig_ep_sf},  Figure~\ref{Fig_tau_ec} plots $e_{\rm p}^{({\rm forced})}$ (from \ref{forced_ecc_general}) against $\tau_{e_{\rm c}}$ (from \ref{eq_tau_ec}) for observed values of $L_{\rm c}/L_{\rm p}$ and $|\mathcal{K}| = e_{\rm p}^{({\rm forced})}/e_{\rm c}$. Evaluation of $\tau_{e_{\rm c}}$ further requires $\gamma_{\rm p}$, which we vary from $10^{-8}$ yr$^{-1}$ to $10^{-6}$ yr$^{-1}$ in making Figure~\ref{Fig_tau_ec}. These values for $\gamma_{\rm p}$ are inspired by our fiducial model which predicts $\tau_a \sim 10^9$--$10^{10}$ yr for $e_{\rm p} \sim 0.03$ (Figure \ref{Fig_inspiral_time}). Many of the observed systems have values for $e_{\rm p}^{({\rm forced})}$ that differ from our fiducial $e_{\rm p} = 0.03$; for these, a more accurate determination of $\tau_{e_{\rm c}}$ would require tailored models, which we defer to future work. In any case, Figure \ref{Fig_tau_ec} argues that some systems can have percent-level eccentricities that can be sustained over Gyr timescales.

\section{Solution method for the Laplace Tidal Equations}\label{LTE_Solution}
We describe the method used to derive the linear system of Equations (\ref{lin_sys_1}) and (\ref{lin_sys_2}). The approach is similar to that in \citet{farhat2022resonant} and \citet{auclair2023can} but we provide it in full here for reference. 

We first establish a useful identity. If one combines the identity
\begin{equation}
   \grad \cdot \left( \Phi \grad \Psi \cross \hat{r} \right) = \left( \grad \Psi \cross \hat{r} \right) \cdot \grad \Phi
\end{equation}
with Gauss's theorem
\begin{equation}
    \int_{\mathcal{O}} \grad \cdot \left( \Phi \grad \Psi \cross \hat{r} \right) d A = \oint_{\partial \mathcal{O}} \Phi \left( \grad \Psi \cross \hat{r} \right) \cdot \hat{n} \, d \ell,
\end{equation}
$dA$ being an infinitesimal area element of the ocean domain $\mathcal{O}$, and $d\ell$ being a length element on the terminator, one obtains
\begin{equation}\label{misc0}
    \int_{\mathcal{O}} \left(  \grad \Phi \right) \cdot \left( \grad \Psi \cross \hat{r} \right)  d A = \oint_{\partial \mathcal{O}} \Phi \left( \grad \Psi \cross \hat{r} \right) \cdot \hat{n} \, d \ell.
\end{equation}
The right hand side of this equation is zero, by virtue of the imposed boundary condition on the stream function. Then the curl-free and divergence-free components of the tidal flow on the left hand side are orthogonal.

Returning to the coupled system of LTEs, substituting the series expansions of $\Phi$ and $\Psi$ in the momentum conservation equation \eqref{momentum1}, and then taking the dot product with $\grad\phi_r$, we get:
\begin{align}\nonumber
    \sum_{s=0}^\infty &\left(\partial_t^2p_s+\sigma_{\rm R}\partial_t p_s +gHp_s\mu_s -g{\zeta}_{\mathrm{eq},s}\right)\grad\phi_s\cdot\grad\phi_r \\\nonumber
    &+\left(\partial_t^2p_{-s}+\sigma_{\rm R}\partial_tp_{-s}\right)\left(\grad\psi_s\cross\hat{r}\right)\cdot\grad\phi_r 
    \\\nonumber&+ \partial_t p_s\left(\Vec{f}\cross\grad\phi_s\right)\cdot\grad\phi_r
    \\&+\partial_t p_{-s}\left[\Vec{f}\cross\left(\grad\psi_s\cross\hat{r}\right)\right]\cdot\grad\phi_r=0.\label{misc1}
\end{align}
Using Green's first identity, we compute the product of the gradient of the two eigenfunctions:
\begin{align}\nonumber
\int_\mathcal{O}\grad\phi_s\cdot\grad\phi_r dA &=\int_{\partial\mathcal{O}} \phi_s\left(\grad\phi_r\cdot\hat{n}\right) d\ell - \int_\mathcal{O}\phi_s\grad^2\phi_rdA\\
&=\mu_r\delta_{r,s}.
\end{align}
The latter equality follows from the first term on the right hand side vanishing due to the boundary condition of the potential function, and using the eigenvalue equation with the normalization condition of the eigenfunctions on the second term.
Integrating \eq{misc1} over the magma ocean domain, we find
\begin{align}\label{pr2}\nonumber
    \sum_{s=0}^\infty &\left(\partial_t^2p_s+\sigma_{\rm R}\partial_t p_s +gHp_s\mu_s -g{\zeta}_{\mathrm{eq},s}\right)\mu_r \delta_{r,s} \\\nonumber
    &+\left(\partial_t^2p_{-s}+\sigma_{\rm R}\partial_tp_{-s}\right) \int_\mathcal{O}\left(\grad\psi_s\cross\hat{r}\right)\cdot\grad\phi_r \,dA \\\nonumber
    &+ \partial_t  p_s \int_\mathcal{O} \Vec{f}\cdot \left(\grad\phi_s\cross\grad\phi_r\right)dA\\
    &+\partial_t p_{-s}\int_\mathcal{O} \left(\Vec{f}\cdot\hat{r}\right) \left(\grad\phi_r\cdot\grad\psi_s\right)dA=0.
\end{align}
The second term vanishes by virtue of the identity we established in \eq{misc0}. Using $\Vec{f}=2\Omega\cos\theta\hat{r}$ we rearrange the equation to obtain:
\begin{align}\label{mscI}\nonumber
 &\left(\partial_t^2p_r+\sigma_{\rm R}\partial_t p_r +gHp_r\mu_r -g{\zeta}_{\mathrm{eq},r}\right)\mu_r \\\nonumber
 &- 2\Omega\sum_{s=1}^\infty \partial_t  p_s \int \cos\theta\hat{r}\cdot \left(\grad\phi_r\cross\grad\phi_s\right)dA\\
&+2\Omega\sum_{s=1}^\infty\partial_t p_{-s}\int \cos\theta \left(\grad\phi_r\cdot\grad\psi_s\right)dA=0.
\end{align}
Going through a similar procedure but starting by taking the dot product of the momentum equation with $\grad\psi_r\cross\hat{r}$ instead of $\grad\phi_r$, we obtain:
\begin{align} \label{mscII} \nonumber
 &\partial_t^2p_{-r}+\sigma_{\rm R}\partial_t p_{-r}- \frac{2\Omega}{\nu_r}\sum_{s=1}^\infty \partial_t  p_s \int \cos\theta \left(\grad\phi_s\cdot\grad\psi_{r}\right)dA\\
&-\frac{2\Omega}{\nu_r}\sum_{s=0}^\infty\partial_t p_{-s}\int \cos\theta \hat{r}\cdot \left(\grad\psi_r\cross\grad\psi_s\right)dA=0. 
\end{align}
Following previous works \citep[][]{proudman1920dynamical1,proudman1920dynamical,longuet1970free,webb1980tides,farhat2022resonant}, we identify in the system of Eqs. \eqref{mscI} and \eqref{mscII} the so-called gyroscopic coefficients:
\begin{align} \nonumber \label{gyro1}
 \beta_{r,s}&=-\int_\mathcal{O} \cos\theta \, \hat{r} \cdot \left( \grad\phi_r\cross\grad\phi_s \right) dA,\\
    \nonumber
     \beta_{r,-s}&= \ \ \, \int_\mathcal{O} \cos\theta \, \grad\phi_r\cdot\grad\psi_s dA,   \\ \nonumber
         \beta_{-r,s}&=-\int_\mathcal{O} \cos\theta \, \grad\psi_r\cdot\grad\phi_s dA,\\
    \beta_{-r,-s}&=-\int_\mathcal{O} \cos\theta \, \hat{r} \cdot \left( \grad\psi_r\cross\grad\psi_s \right) dA,
\end{align}
with $\beta_{-s,r}= -\beta_{r,-s}$. With these definitions, we recast the system of Eqs. \eqref{mscI} and \eqref{mscII} into:
\begin{equation} 
\label{system_1_before}
    \partial_t^2p_r+\sigma_{\rm R}\partial_t p_r +gH\mu_rp_r -g{\zeta}_{\mathrm{eq},r} + \frac{2\Omega}{\mu_r}\sum_{s=-\infty}^{s=\infty}\beta_{r,s}\partial_tp_s =0,
\end{equation}
\begin{equation}
\label{system_2_before}
    \partial_t^2p_{-r} +\sigma_{\rm R}\partial_tp_{-r} +\frac{2\Omega}{\nu_r}\sum_{s=-\infty}^{s=\infty}\beta_{-r,s}\partial_t p_s =0.
\end{equation}
Transforming this system to the Fourier domain using the tidal forcing frequency $\sigma$, i.e. $\partial_t=i\sigma$, we recover Eqs. (\ref{lin_sys_1})-(\ref{lin_sys_2}) in the main text.

\section{Hansen Coefficients}\label{Appendix_Hansen_coeffs}
We compute the Hansen coefficients $X^{n,m}_k(e_{\rm p})$ as the Fourier coefficients in the mean anomaly $M_{\rm p}$ of the $2\pi$-periodic function
\begin{equation}
    f(M_{\rm p}) \equiv \left(\tfrac{r_{\rm p}}{a_{\rm p}}\right)^{n} e^{\,i m v_{\rm p}(M_{\rm p})},
\end{equation}
namely
\begin{equation}
   X^{n,m}_k(e_{\rm p}) = \frac{1}{2\pi}\!\int_{0}^{2\pi}\!\left(\frac{r_{\rm p}}{a_{\rm p}}\right)^{n} e^{\,i m v_{\rm p}(M_{\rm p})} e^{-i k M_{\rm p}}\,\mathrm{d}M_{\rm p},
\end{equation}
or equivalently,
\begin{equation}
    \left(\frac{r_{\rm p}}{a_{\rm p}}\right)^{n} e^{\,i m v_{\rm p}} = \sum_{k=-\infty}^{+\infty} X^{n,m}_k(e_{\rm p})\,e^{\,i k M_{\rm p}},
\end{equation}
with the Kepler relations $M_{\rm p} = E_{\rm p} - e_{\rm p}\sin E_{\rm p}$,  ${r_{\rm p}}/{a_{\rm p}} = 1 - e_{\rm p}\cos E_{\rm p}$, and
$v_{\rm p} = \operatorname{atan2}\!\bigl(\sqrt{1-e_{\rm p}^{2}}\sin E_{\rm p},\,\cos E_{\rm p} - e_{\rm p}\bigr)$.
Practically, we sample $M_{{\rm p},j} = j\,\Delta M_{\rm p}$ uniformly on $[0,2\pi]$ with $N=2^{p}$ points (so $\Delta M_{\rm p}=2\pi/N$), solve Kepler’s equation for $E_{{\rm p},j}$, form $f_{{\rm p},j}=(1-e_{\rm p}\cos E_{{\rm p},j})^{n}\,e^{\,i m v_{{\rm p},j}}$, and evaluate the discrete Fourier transform
\begin{equation}
Z_k \;=\; \frac{1}{N}\sum_{j=0}^{N-1} f_{{\rm p},j}\,e^{-i k M_{{\rm p},j}}.
\end{equation}
For elliptic motion the coefficients $X^{n,m}_k(e_{\rm p})$ are real, so numerically we set $X^{n,m}_k(e_{\rm p})=\Re\{Z_k\}$.
This FFT implementation is consistent with the classical definition and properties of Hansen coefficients \citep{hughes1981computation}, their generalized connection to Laplace coefficients \citep{laskar2005note}, and modern disturbing-function expansions with small $e_{\rm p}$ scalings \citep{mardling2013new}.

\section{Mantle phases}\label{Appendix_mantle_phase_diagram}
We first structure the phase diagram by adopting parameterized cubic spline radial functions for the solidus, $T^{\rm sld}(r)$, the liquidus, $T^{\rm liq}(r)$, and the temperature profile $T^{\rm crit}(r)$ corresponding to a critical melt fraction   $F_{\rm m}=40\%$  \citep[][]{zhang1994melting,fiquet2010melting,miyazaki2019timescalea}{}{}. For temperatures above (below) $T^{\rm liq}$ ($T^{\rm sld}$), the silicate mantle layer is fully molten (solidified), while the $T^{\rm crit}$ profile in between defines the rheological transition from a deformable solid matrix that is penetrated by a network of percolating melt to a fluid magma layer with suspended solids, and around which a regime of mushy magma exists \citep[e.g.][]{abe1997thermal}{}{}.

To construct the mantle adiabats living in this phase diagram, we integrate the standard entropy conserving gradient for a chosen mantle potential temperature $T_{\rm p}$ and a reference pressure $P_{0}$ and following it downward:
\begin{equation}\label{adiabatic_profile}
  \frac{{\rm d}T}{{\rm d}P}\Bigg|^{\rm ad} = \frac{\alpha(P,T)\,T}{\rho(P,T)\,C_{\rm p}(P,T)} ,
\end{equation}
where $\alpha$ is thermal expansivity, $\rho$ is the density, and $C_{\rm p}$ is the isobaric heat capacity. These material properties, along with the phase diagram functions, are evaluated on the fly using the cubic spline fits and parametrized relations delineated explicitly in \citet{miyazaki2019timescalea} and \citet{korenaga2023rapid}. The only modification we implement is that, while the relations therein are Earth depth dependent, we move to a pressure dependent scale for an arbitrary $(R_{\rm p},M_{\rm p})$ by assuming hydrostatic balance and rescaling the outer boundary gravity.  We assume that the mantle volume of each planet is proportionally similar to that of the Earth \citep[$\sim 84\%$ of the planetary volume, e.g.~][]{stacey2008physics}{}{}, which allows us to define the pressure base at the core-mantle boundary.

In the biphasic regime where mantle temperatures lie between the solidus and the liquidus, the adiabatic profile  of \eq{adiabatic_profile} is modified to include the latent heat of fusion à la Eq.~(39) of \citet{farhat2025tides}:
\begin{equation}\label{mantle_adiabat2}
    \frac{dT}{dP}\Bigg|_{T^{\rm sld} >T>T^{\rm liq}}=  \frac{dT}{dP}\Bigg|^{\rm ad}  + \frac{H_{\rm f}}{C_{\rm p}}\gamma,  
\end{equation}
with $H_{\rm f}$ denoting the latent heat of fusion, linearly interpolated from $6\times10^5$~J~kg$^{-1}$ at the surface to
$9 \times10^6$~J~kg$^{-1}$ at the core-mantle boundary, and $\gamma$ defining the gradient of melt with pressure  \citep[][]{langmuir1992petrological}{}{}:
\begin{equation}
    \gamma = \left({\frac{dT}{dP}\Bigg|^{\rm ad}-\frac{dT^{\rm sld}}{dP}}\right)\left({\frac{H_{\rm f}}{C_{\rm p}} + \frac{dT}{dF_{\rm m}}}\right)^{-1}.
\end{equation}

Finally, we define for the mantle a radial viscosity profile that follows the Arrhenius equation, only modified to include the effect of partial melt within the solid matrix, i.e.,

\begin{equation}\label{viscosity_temperature}
    \eta(r,T) = \eta_{\rm s}(r)\exp\{-B F_{\rm m}\} \exp\left\{\frac{E_{\rm a}}{R T^{ s}(r)}\left(\frac{T^{ s}(r)}{T}-1\right) \right\}.
\end{equation}
Here $\eta_{\rm s}$ is the viscosity profile of the solid background ignoring the effect of the melt, i.e., at the solidus temperature; $B$ is an experimentally fitted parameter that ranges between 10 and 40 \citep[e.g., $B\sim 26$ for diffusion creep; ][]{mei2002influence}{}{}, $E_{\rm a}=300 \ {\rm  kJ \ mol^{-1}}$ is the activation energy, and $R= 8.314462 \ {\rm J \ mol^{-1} \  K^{-1}}$ is the ideal gas constant \citep[][]{korenaga2006archean,tiesinga2021codata}{}{}. 

\section{Tidal Heating in the Solid Viscoelastic mantle}\label{Appendix_solid_tides}
We consider the solid part of the mantle to be incompressible, with viscosity $\eta_{\rm s}$ and shear modulus $\mu_{\rm s}$, and we adopt a Maxwell rheology to describe its strain under tidal stresses \citep[e.g.][]{efroimsky2012tidal,bagheri2019tidal}. We set $\mu_{\rm s}=75~{\rm GPa}$ \citep[e.g.][]{tobie2005tidal}. The Maxwell time and dimensionless frequency are $\tau_{\rm M}=\eta_{\rm s}/\mu_{\rm s}$ and $x=|\sigma|\tau_{\rm M}.$ The complex shear modulus is
\begin{equation}
    \hat{\mu}_{\rm s}(\sigma) =\mu_{\rm s}\frac{i\,x}{1+i\,x}.
\end{equation}
The degree-n tidal  Love numbers associated with the solid interior is given by \citep[e.g.][]{munk1960continentality}
\begin{equation}
    k_n(\sigma)= \frac{3}{2(n-1)}\frac{1}{1+\tilde{\mu}_n(\sigma)},
\end{equation}
where we have introduced the dimensionless effective shear modulus $\tilde{\mu}_n(\sigma) = A_n \hat{\mu}_{\rm s}(\sigma) / {\mu}_{\rm s}$, with the gravitational coupling $A_n$ captured by
\begin{equation}
    A_n = \frac{4(2n^2+ 4n + 3) \pi R_{\rm s}^4 \mu_{\rm s}}{3nGM_{\rm s}},
\end{equation}
with $R_{\rm s}$ and $M_{\rm s}$ denoting the radius and mass of the solid part of the mantle, respectively. The imaginary part of the Love number, governing dissipation, is then
\begin{equation}
\Im\left\{{k}_n(\sigma)\right\}
=-\,\frac{3}{2(n-1)}\,
\frac{A_n\,x}{1+\bigl(1+A_n\bigr)^2 x^2}.
\end{equation}
Tidal heating in the solid then follows
\begin{equation}
\mathcal{P}_{\rm T;s}
=\frac{R_p}{8\pi G}\sum_{s=-\infty}^{\infty}
\sum_{n=2}^{\infty}\sum_{m=0}^{n}(2n+1)\sigma_{m}^{s}\bigl|U^{T;s}_{n,m}\bigr|^{2}\Im\left\{{k}_n\bigl(\sigma)\right\},
\end{equation}
with $U^{T;s}_{n,m}$ given by \eq{Upqs}.
\end{document}